\documentclass[english,aps,twocolumn,showpacs,amsmath,amssymb,prb,footinbib]{revtex4-2}
\usepackage[T1]{fontenc}
\usepackage[latin9]{inputenc}
\usepackage{amsmath}
\usepackage{graphicx}
\usepackage{esint}

\makeatletter
\usepackage{graphicx}
\usepackage{color}
\usepackage{amsmath}
\usepackage{psfrag}
\usepackage[abs]{overpic}
\usepackage{enumitem}
\usepackage{natbib}
\usepackage{float}
\usepackage{verbatim}
\usepackage[%
  colorlinks=true,
  urlcolor=blue,
  linkcolor=blue,
  citecolor=blue
]{hyperref}

\makeatother

\usepackage{babel}
\begin{document}
\title{Circuit quantization with time-dependent magnetic fields for
realistic geometries }
\author{R.-P. Riwar$^1$ and D. P. DiVincenzo$^{2,1}$}
\affiliation{$^1$ Peter Gr\"unberg Institute, Theoretical Nanoelectronics, D-52425 J\"ulich, Germany and J\"ulich-Aachen Research Alliance (JARA), Fundamentals of Future Information Technologies,
D-52425 J\"ulich, Germany \\ $^2$ Institute for Quantum Information,
RWTH Aachen University, D-52056 Aachen, Germany}

\begin{abstract}
Quantum circuit theory has become a powerful and indispensable tool
to predict the dynamics of superconducting circuits. Surprisingly
however, the question of how to properly account for a time-dependent
driving via external magnetic fields has hardly been addressed
so far. Here, we derive a general recipe to construct a low-energy
Hamiltonian, taking as input only the circuit geometry and the solution
of the external magnetic fields. 
A gauge fixing procedure for the scalar and vector potentials is given which assures that time-varying magnetic fluxes make contributions only to the potential function in the Schr{\"o}dinger equation.
Our proposed procedure is valid for
continuum geometries and thus significantly generalizes previous efforts,
which were based on discrete circuits. We study some implications
of our results for the concrete example of a parallel-plate SQUID circuit. We show that if we insist on representing 
the response of this SQUID with individual, discrete capacitances associated with each individual Josephson junction, this is only possible if we permit the individual capacitance values
to be negative, time-dependent or even momentarily singular.
Finally, we provide some experimentally testable predictions, such
as a strong enhancement of the qubit relaxation rates arising from the effective negative
capacitances, and the emergence of a Berry phase due to time dependence of these
capacitances.
\end{abstract}
\maketitle

\section{Introduction}

Superconducting circuits have proved to be a successful design kit for the creation of new quantum systems, especially for quantum information processing. Artificial atoms, and crafted light-matter couplings, have emulated and extended nature, using the new paradigm of {\em circuit quantum electrodynamics} (cQED)~\cite{blais2020circuit}. This paradigm has now produced quantum-computing devices of unrivalled complexity~\footnote{See \href{ https://www.ibm.com/blogs/research/2020/09/ibm-quantum-roadmap/}{https://www.ibm.com/blogs/research/2020/09/ibm-quantum-roadmap/}}.  cQED has been underpinned by a very handy scheme for turning the description of an electric circuit into a quantum Hamiltonian~\cite{Yurke_1984,Devoret_1997,Burkard_2004,Ulrich_2016,Vool_2017}; taking a lumped-element point of view, one identifies capacitors, inductors, and Josephson junctions, each of which contributes its own piece to this Hamiltonian. It has always been understood that, as successors of the superconducting quantum interference device (SQUID), these circuits have a very effective means for in-place control of their quantum characteristics, via the setting of threading magnetic fields.

However, it was very recently remarked by You, Sauls, and Koch~\citep{You2019} that some important aspects of the Hamiltonian description of cQED
have not yet been appropriately discussed, when including time-dependent
(electro)magnetic fields. The problem is in fact of a very general nature.
When describing a quantum system with a Hamiltonian $H\left(x\right)$,
depending on an external parameter $x$ (in our case, the magnetic
flux), there is an ambiguity when including a time-dependent drive
$x\rightarrow x\left(t\right)$: the system dynamics differ, depending
on the choice of basis through the unitary operator $U$, by the extra term in the Hamiltonian operator in the Schr\"odinger
equation, $-i\dot{x}U\partial_{x}U^{\dagger}$. For a general quantum
system, finding the suitable basis may be challenging, since a microscopic
derivation of $H$ is not always available. In \citep{You2019}, this basis was found for lumped-element quantum circuits by means
of an ``irrotational'' gauge for the Hamiltonian, $H_{\text{irr}}$,
where the extra term related to $U$ vanishes.

In the present work, we strongly generalize the notion of an irrotational
gauge to continuous circuit models. Such a generalization is, in spirit,
reminiscent of efforts to extend the single-mode description of the
circuit environment \citep{Houck_2008,Bourassa_2009,Niemczyk_2010,Filipp_2011,Viehmann_2011}
to more realistic models with continuous degrees of freedom \citep{Nigg_2012,Firat_2014}.
In particular, we find that the suitable irrotational gauge for the
vector potential, $\mathbf{A}_{\text{irr}}(\mathbf{r},t)$, is the Coulomb gauge,
supplemented with an additional boundary condition at the (super)conductor surfaces. The part of $\mathbf{A}_{\text{irr}}$ parallel to the surfaces connects to the London gauge (and thus to the Meissner screening currents), whereas a generally dominant perpendicular part relates to surface charges generated by the time-varying magnetic field. With this gauge
we are able to formulate a precise set of steps in order to arrive
at the correct time-dependent Hamiltonian, starting from a general device geometry and a given time dependent magnetic field as the input.

In addition, our proposed gauge allows for a considerable reduction of the computational
effort in experimentally relevant device geometries. Namely, due to the widespread use of the Niemeyer-Dolan technique \citep{Niemeyer_1976,Dolan_1977},
Josephson junctions 
are often incorporated in thin, long filamentary structures. The conditions of the irrotational gauge allow us to show that if
such junction filaments are sufficiently thin, they can be eliminated
from the device geometry altogether when we solve for $\mathbf{A}_{\text{irr}}$.

We illustrate our results with the example of a 1D SQUID-circuit geometry where
analytic solutions are accessible. Importantly, while a mapping from the continuous
circuit model onto a lumped-element circuit still works, it comes
in general at the expense of having to assign negative or even time-dependent
capacitances to each Josephson junction, depending on the circuit
geometry and magnetic field distribution. We note though that the
effect here is of purely dynamical origin, and has nothing to do with
any material properties such as, e.g., negative capacitances in ferroelectric
materials \citep{Landauer_1976,Catalan_2015,Ng_2017,Hoffmann_2018,Lukyanchuk_2019,Hoffmann_2020}.
In particular, the total capacitance of an island (sum of junction
and other capacitances) remains constant and positive. Nonetheless,
negative and time-dependent capacitances lead to observable consequences.
The former results in significant deviations for the prediction of
qubit relaxation rates, whereas the latter yields a finite Berry phase
in the adiabatic regime, the measurement of which is experimentally
achievable \citep{Abdumalikov_2013,Roushan_2014}.

For a general circuit model, finding $\mathbf{A}_{\text{irr}}$ in
practice requires finding the low-frequency solution of the electric
field $\mathbf{E}_{\dot{B}}(\mathbf{r},t)$ as a response to the time-dependent
magnetic field $\mathbf{B}(\mathbf{r},t)$. This is essentially equivalent
to a first order expansion in the finite-frequency circuit impedance
calculation Feynman presented in his 23\textsuperscript{rd} lecture
\citep{Feynman_1963}. As a result, we expect that state-of-the-art
numerical field solvers for capacitances and inductances (such as FastCap\textsuperscript{TM}
and FastHenry\textsuperscript{TM}, built on the fundamental methodological insights of Ruehli~\cite{5391464,1127927}), often deployed for realistic
circuit geometries, would require an extension with corrections due to finite driving frequencies. 

Finally, we expect that our results are relevant also for the time-dependent
control of topological circuits. Namely, to guarantee for a certain
topological system in a given symmetry class \citep{Altland_1997,Chiu_2016}
to remain in said class under the influence of an external drive imposes
in general the same symmetry restrictions onto the extra term $-i\dot{x}U\partial_{x}U^{\dagger}$.
In this regard, we believe that our work could for instance be of
importance to reevaluate the topological protection of time-dependent
flux-driving of Majorana-based circuits \citep{van_Heck_2012} for
realistic device geometries.

This paper is structured as follows. In Sec.~\ref{sec:general-geometries}
we develop the general recipe to construct the Hamiltonian for quantum
circuits in the presence of a time-dependent drive, including the
definition of a general notion of an irrotational gauge for the vector
field. We apply these principles at the example of a simple SQUID
geometry in Sec.~\ref{sec:anomalous-capacitances} and discuss measurable
consequences of anomalous capacitances. In Sec.~\ref{sec:refined-approach}
we establish a connection between our work and Ref.~\citep{You2019}
through refining the lumped-element approach and going to a continuum
limit. The conclusions are presented in Sec.~\ref{sec:conclusions}.

\begin{figure}
\centering\includegraphics[width=0.8\columnwidth]{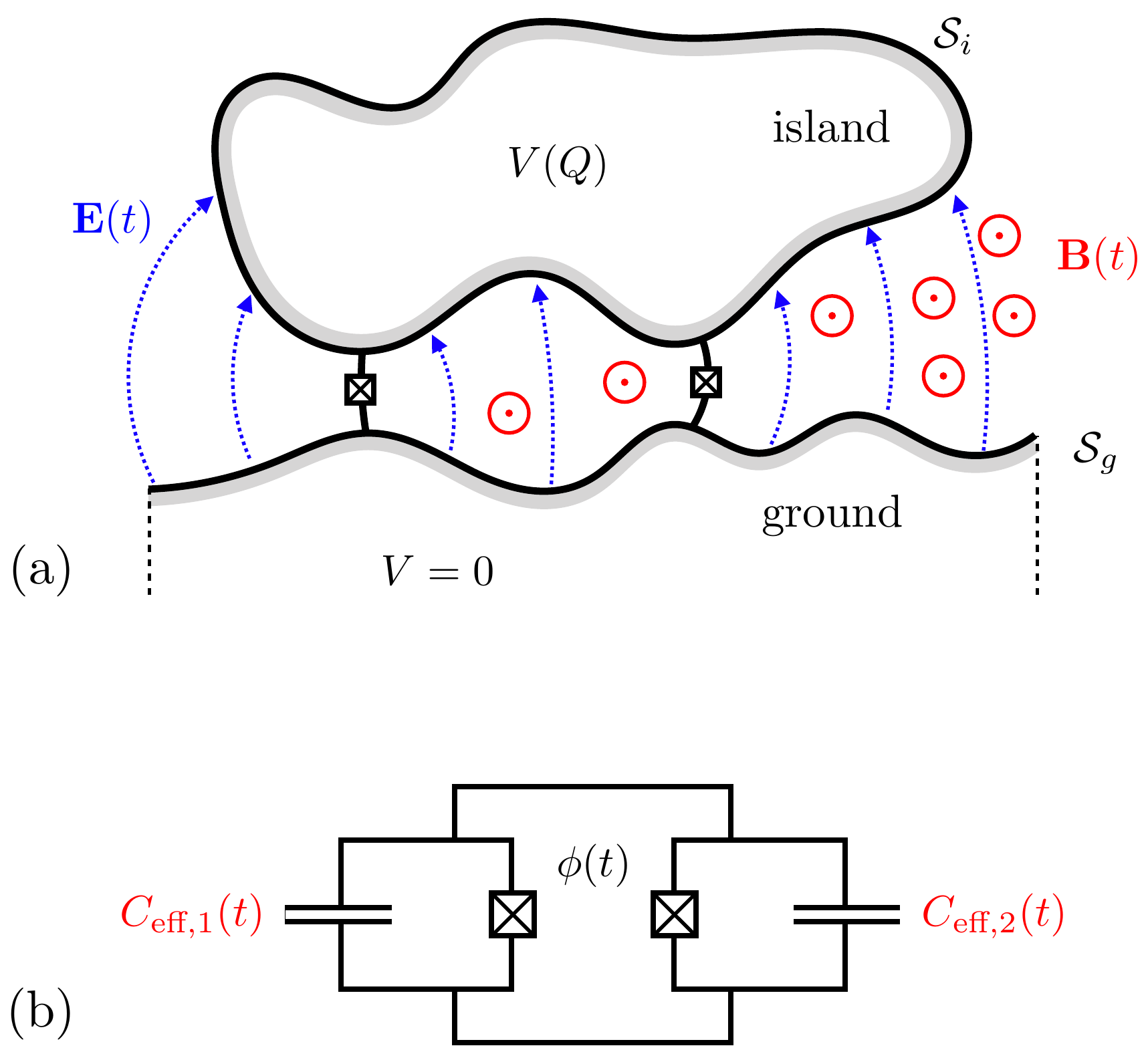}

\caption{Some of the main accomplishments of this work. (a) We develop a recipe to derive
the correct time-dependent Hamiltonian for circuits with a general
geometry, taking as input the time-dependent magnetic field $\mathbf{B}\left(t\right)$ (red), from which the time-dependent electric field $\mathbf{E}\left(t\right)$ (blue) is derived. (b) We find that in this general case, it is not always possible
to assign well-behaved capacitances to the two Josephson junction
of a SQUID. While the total capacitance $C_{\text{tot}}=C_{\text{eff,}1}\left(t\right)+C_{\text{eff},2}\left(t\right)$
is guaranteed to be constant in time and positive definite, if we force a
mapping to a model with effective junction capacitances $C_{\text{eff},j}\left(t\right)$, these may be negative, time-dependent or even momentarily singular.\label{fig_intro}}
\end{figure}

\section{General geometries and gauge fixing\label{sec:general-geometries}}

For visualization purposes, throughout this section we refer to the
example of a SQUID circuit with a generic geometry (Fig.~\ref{fig_intro}(a)),
driven with a time-dependent magnetic field $\mathbf{B}\left(\mathbf{r},t\right)$.
We comment at the end of this section, how to generalize to arbitrary
circuits. We assume that the lumps making up the circuit are perfect
(super)conductors, expelling both electric and magnetic fields from
their interior. In the SQUID, we refer to one of the lumps as the
island $i$ and the other one as the ground $g$ (setting its voltage
to zero). They have surfaces $\mathcal{S}_{l}$, with
vector $\mathbf{n}_{l}$ normal to the surface. The index $l=i,g$
enumerates the lumps. Only the surface of the island has to form a
closed manifold, the ground surface can be (and usually is) extended.
Two Josephson junctions connecting the two lumps enclose a finite
area $\mathcal{A}$, which is pierced by the flux
\begin{equation}
\Phi\left(t\right)=\iint_{\mathcal{A}}d^2x\,\,{\bf n}_\mathcal{A}\cdot{\bf B}\left({\bf r},t\right),\label{eq:flux}
\end{equation}
where $\mathcal{A}$ will lie in the $\left(x,y\right)$-plane in our examples later.

\subsection{The state of the art\label{the_state_of_the_art}}

Circuits as in Fig.~\ref{fig_intro}(a) are usually right away transformed
into a lumped element circuit as shown in Fig.~\ref{fig_intro}(b),
where to each junction with Josephson energy $E_{Jk}$ one assigns
a capacitance $C_{k}$.

Any external flux $\Phi(t)$ enters the Hamiltonian as a phase $\phi\left(t\right)=2\pi\Phi\left(t\right)/\Phi_{0}$
where $\Phi_{0}=h/\left(2e\right)$ is the flux quantum. For a time-independent
flux, $\dot{\phi}=0$, the dynamics of the SQUID circuit is described
by a Hamiltonian of the form
\begin{multline}
\widehat{H}_{\alpha}=\frac{\left(2e\right)^{2}}{2C_{\text{tot}}}\widehat{n}^{2}-E_{J1}\cos\left(\widehat{\varphi}+\alpha\phi\right)\\
-E_{J2}\cos\left(\widehat{\varphi}+\left[\alpha-1\right]\phi\right),\label{eq:H_alpha}
\end{multline}
where the island charge and phase operators satisfy the commutation
relations $\left[\widehat{\varphi},\widehat{n}\right]=i$, and the
total capacitance of the island is $C_{\text{tot}}=\sum_{k=1,2}C_{k}$.
Importantly, in the absence of time-dependent driving, the constant,
real parameter $\alpha$ can be chosen arbitrarily: all Hamiltonians
$H_{\alpha}$ will give rise to the same predictions irrespective
of $\alpha$, making $\alpha$ a gauge degree of freedom, expressed
by the unitary transformation $\widehat{H}_{\alpha}=\widehat{U}_{\alpha}\widehat{H}_{0}\widehat{U}_{\alpha}^{\dagger}$,
with $\widehat{U}_\alpha=e^{i\alpha\phi\widehat{n}}$.

When including a time-dependent driving of $\mathbf{B}$, resulting
in a time-dependent flux $\phi$, it seems tempting to simply take
the Hamiltonian (\ref{eq:H_alpha}) for a given $\alpha$, and insert a time-dependent
flux $\phi\rightarrow\phi\left(t\right)$. However, as discussed in
Ref.~\citep{You2019}, this is generally not correct. For a time-dependent $\phi\left(t\right)$,
Hamiltonians (\ref{eq:H_alpha}) with different $\alpha$ give rise to a physically distinct
time evolution, simply because the unitary transformation $\widehat{U}_{\alpha}$
is now time dependent, and therefore needs to show up in the Schr\"odinger equation
through the extra term $-i\widehat{U}_{\alpha}\partial_{t}\widehat{U}_{\alpha}^{\dagger}=-\alpha\dot{\phi}\widehat{n}$.
Hence, to correctly predict the dynamics of the time-dependent problem,
$\alpha$ has to be fixed. The authors of Ref. \citep{You2019} deploy
the Lagrangian method including time-dependent contraints, and show
that for the SQUID, the time dependent version of the Hamiltonian Eq.~(\ref{eq:H_alpha}) is correct when 
\begin{equation}
\alpha=\frac{C_{2}}{C_{\text{tot}}}.\label{eq:alpha_C}
\end{equation}
The authors refer to this as the irrotational gauge. For any other
gauge, there will appear linear terms $\sim\dot{\phi}\widehat{n}$
in the Hamiltonian, due to the time-dependent basis transformation.
To summarise, \citep{You2019} argues that a Hamiltonian of the form
of Eq.~(\ref{eq:H_alpha}) can only be found for the time-dependent
problem when fixing $\alpha$ through the knowledge of the capacitances
of the Josephson junctions $C_{k}$. 

Crucially, while we deem the procedure proposed in Ref.~\citep{You2019}
to be correct when based on the assumptions stated therein, it comes with
one particular assumption which we want to put under further scrutiny:
the hypothesis that one can assign a definite, constant, positive capacitance
$C_{k}$ to each junction. 

In the following, we will consider the
problem from a pure electrodynamic point of view, with general, realistic
device geometries living in continuous 3D space. Importantly, we
will show that indeed, in general this critical hypothesis must
be relaxed to some extent. In particular, we find that a mapping of
the realistic circuit Fig.~\ref{fig_intro}(a) to the lumped element circuit
in Fig.~\ref{fig_intro}(b) is only possible if one accepts potentially
negative, time-dependent, or even momentarily singular capacitances $C_{k}\left(t\right)$,
which depend not only on the circuit geometry, but also on the spatial
distribution of the magnetic field. It is only guaranteed that the
total capacitance $C_{\text{tot}}=\sum_{k}C_{k}\left(t\right)$ remains
constant and positive.

The demonstration of these facts will involve the following two steps.
We first lay out the precise conditions under which the induced electric
field $\mathbf{E}\left({\bf r},t\right)$ can be uniquely determined based
on a certain circuit geometry as well as a given magnetic field $\mathbf{B}\left({\bf r},t\right)$.
Then, we discuss how we can, given $\mathbf{E}({\bf r},t)$, arrive at the scalar and vector potentials with the gauge fixed according to a generalization of the irrotational condition of~\cite{You2019}.

\subsection{Solutions of time-dependent electromagnetic fields\label{unique_fields}}

Let us assume that a specific magnetic field texture $\mathbf{B}\left({\bf r},t\right)$ is
provided by some external source. Because of the Meissner effect,
the magnetic field is expelled from the inside of the conductors, on the length scale given by the London penetration depth $\lambda$. In order to figure out the resulting electric field $\mathbf{E}$, it is useful to separate the discussion for the field inside the conductors and outside.

First, let us discuss the interior of the conductors. The Meissner effect involves a screening supercurrent, given by Amp\`{e}re's law $\mathbf{j}=\nabla \times \mathbf{B}/\mu_0$. For a time-varying magnetic field, there immediately emerges the resulting electric field due to the first London equation, $\mathbf{E}=\mu_0 \lambda^2 \partial_t\mathbf{j}$. 

Let us say a couple of words about the nature of this electric field. Due to current conservation $\nabla\cdot\mathbf{j}=0$, so the current, and therefore the electric field inside, must be fully tangential to the surface, $\mathbf{n}_l \cdot\mathbf{E}=0$. Just like the magnetic field, this tangential electric field penetrates the superconductor to a distance $\lambda$ inside the surface, as determined by the ac Meissner screening currents. This insight is of importance for the now following discussion of the exterior electric fields.

Namely, for said exterior, the resulting
electric field $\mathbf{E}\left({\bf r},t\right)$
is determined by
\begin{align}
\nabla\times\mathbf{E} & =-\dot{\mathbf{B}},\label{eq_rot_E}\\
\nabla\cdot\mathbf{E} & =0 \ .\label{divfree}
\end{align}
As mentioned in the introduction, this is
equivalent to a first order expansion of the field solutions discussed
by Feynman \citep{Feynman_1963}. 
Moreover, the electric field at the conductor surfaces has to satisfy the general boundary
condition~\cite{Jackson_book}, $\mathbf{n}_l\times(\mathbf{E}_1-\mathbf{E}_2)=0$, where $\mathbf{E}_{1,2}$ are the electric fields at the interface, when approaching it from the outside or from the inside, respectively. Here, this amounts to
\begin{equation}
\mathbf{n}_{l}\times\mathbf{E}\biggr|_{\mathbf{x}\in\mathcal{S}_{l}}=\lambda^2\mathbf{n}_l\times\left(\nabla\times \dot{\mathbf{B}}\right)\biggr|_{\mathbf{x}\in\mathcal{S}_{l}},\label{eq_E_boundary}
\end{equation}
where $\mathbf{E}$ on the left-hand side is the solution, when approaching the surface $\mathcal{S}_l$ from the exterior. The above condition essentially requires the part of the electric field tangential to the conductor surfaces to be continuous, whereas in the exterior, there is an additional component of the electric field, which is allowed to be normal to the surface (henceforth referred to as longitudinal), which abruptly (discontinuously) goes to zero in the interior. Thus, while the tangential part is associated to screening currents, the longitudinal part gives rise to surface charges, localized on the Thomas-Fermi screening length scale, which can be essentially taken to be zero. This approximation, valid for driving frequencies below the plasma frequency, is standard in the literature~\cite{London_book,Bardeen_1955,Bardeen_1957}.  
The solution for $\mathbf{E}$ is
then fixed and unique if we take the island to have a well-defined total charge,
\begin{equation}
\oiint_{\mathcal{S}_{i}}d^{2}x\,\,\sigma_{i}=Q\label{eq_island_charge}
\end{equation}
which is determined through the integral over the above mentioned surface charge
\begin{equation}\label{eq_surface_charge}
\sigma_{i}=\epsilon\left.\mathbf{n}_{i}\cdot\mathbf{E}\right|_{\mathbf{x}\in\mathcal{S}_{i}}.
\end{equation}
From the uniqueness of $\mathbf{E}$, it follows that one can uniquely
decompose the electric field into two components
\begin{equation}
\mathbf{E}=\mathbf{E}_{Q}+\mathbf{E}_{\dot{B}}\label{eq_decomposition}
\end{equation}
where $\mathbf{E}_{Q}$  satisfies Eqs.~(\ref{eq_rot_E}-\ref{eq_island_charge}) for $\dot{\mathbf{B}}=0$, and $\mathbf{E}_{\dot{B}}$ satisfies Eqs.~(\ref{eq_rot_E}-\ref{eq_island_charge}) for $Q=0$. Due to linearity, the total electric field is just the sum of these two contributions.

These two fields have the following properties. Due to a simple scaling argument,
we know immediately that $\mathbf{E}_{Q}$ must be linear in $Q$,
\begin{equation}
\mathbf{E}_{Q}=Q\mathbf{e}\label{eq_E_field_scaling}
\end{equation}
where the vector field $\mathbf{e}$, with units of Newtons per Coulomb squared,
solves the above equations for $\dot{\mathbf{B}}=0$ and $Q=1$.

As for $\mathbf{E}_{\dot{B}}$, it can in the most general case only
be formally represented as a functional 
\begin{equation}
\mathbf{E}_{\dot{B}}=\mathbf{F}\left[\dot{\mathbf{B}}\right].\label{eq:E_functional}
\end{equation}
However, the functional has certain general properties. Firstly, it
only contains integrals over spatial compontents, whereas the time-dependence
is parametrical (because the problem is time local). Therefore, as
long as the circuit geometry stays stationary, $\partial_{t}\left(\mathbf{F}\left[\mathbf{B}\right]\right)=\mathbf{F}\left[\dot{\mathbf{B}}\right]$.
Secondly, due to the linearity of the differential equations, the
functional is likewise linear, $\mathbf{F}\left[a\mathbf{B}\right]=a\mathbf{F}\left[\mathbf{B}\right]$.

Finally, we note that it is in general possible to considerably simplify the boundary condition, Eq.~\eqref{eq_E_boundary}. Namely, if one considers geometries, where the dominant length scales are much larger than the London penetration depth, one is entitled to set $\lambda\rightarrow 0$, leading to,
\begin{equation}
\mathbf{n}_{l}\times\mathbf{E}\biggr|_{\mathbf{x}\in\mathcal{S}_{l}}=0.\label{eq_E_boundary_simple}
\end{equation}
In some sense, this corresponds to rescaling the device geometry to length scales where the London penetration depth is negligibly small. We note that there are model systems with an artificially high degree of radial symmetry (such as, e.g., spherical conductors in a uniform magnetic field, see Ref.~\cite{Matute_1999}) where the longitudinal electric fields must vanish, such that the simplification in Eq.~\eqref{eq_E_boundary_simple} may fail. For generic device geometries however, the above approximation can be expected to work. Later, in Sec.~\ref{sec:anomalous-capacitances} we will show in particular the example of a parallel plate capacitor, where the validity of Eq.~\eqref{eq_E_boundary_simple} can be explicitly shown, when the capacitor dimensions (in particular the plate separation) are larger than $\lambda$. 

\subsection{Irrotational gauge\label{irrotational_gauge}}

In order to formulate a suitable Lagrangian, and then a Hamiltonian,
we have to find the scalar and vector potentials, $V$ and $\mathbf{A}$, which generate the force fields. Starting from the electric field determined
along the lines above, we have to find a solution for
\begin{equation}
\mathbf{E}=-{\boldsymbol{\nabla}} V-\dot{\mathbf{A}}.
\end{equation}
To this equation, there is of course no unique solution - a suitable
gauge has to be chosen. In the same spirit as Ref.~\citep{You2019},
we here want to formulate an irrotational gauge for this generic problem
based on $\mathbf{B}$ and $\mathbf{E}$.

As introduced in Sec. \ref{the_state_of_the_art}, the unique trait
of the irrotational gauge is that there occur no linear terms $\sim\widehat{n}$
in the Hamiltonian, but rather only quadratic terms $\sim\widehat{n}^{2}$. This means that on the level of the Lagrangian, the kinetic energy associated with the voltage difference
between ground and island should likewise be purely quadratic in $V$.
This will be guaranteed by defining the voltage in our irrotational
gauge as
\begin{equation}
V_{\text{irr}}=\int_{\mathcal{L}_{g\rightarrow i}}d\mathbf{l}\cdot\mathbf{E}_{Q},
\end{equation}
where $\mathcal{L}_{g\rightarrow i}$ denotes an arbitrary path from
a point inside $g$ to a point inside $i$. Since $\mathbf{E}_{Q}$
is curl-free, the voltage is thus single-valued as it should be. Moreover,
due to $\mathbf{E}_{Q}=0$ inside the conductors (the ac field penetrating the superconductors on the length scale $\lambda$ is fully included in $\mathbf{E}_{\dot{B}}$), for this gauge choice the voltage
assumes a single, constant value within each conductor volume, such
that the voltage difference between ground and island is well defined,
independent of the starting and end points of $\mathcal{L}_{g\rightarrow i}$~\footnote{In principle, we would be free to choose a gauge where $V$ is not constant within a single bulk of superconductor. This would however artificially and unnecessarily complicate the problem, as the corresponding superconductor could no longer be described by a single phase.}.
Because of the linear dependence of $\mathbf{E}_{Q}$ on $Q$, see
Eq.~(\ref{eq_E_field_scaling}), the irrotational voltage is related
to the island charge $Q$ as
\begin{equation}\label{eq_Q_vs_V_irr}
Q=C_{\text{tot}}V_{\text{irr}},
\end{equation}
where the total capacitance between island and ground is defined as
\begin{equation}
C_{\text{tot}}=\frac{1}{\int_{\mathcal{L}_{g\rightarrow i}}d\mathbf{l}\cdot\mathbf{e}}.\label{eq:capacitance_def}
\end{equation}
Importantly, we see now that for this gauge choice, the kinetic energy
(defined as the time integral of the electric power~\cite{citeulike:9606973}, see also Appendix~\ref{app_T_and_general_gauge}) results in a
purely quadratic term in $V$,
\begin{equation}\label{eq_T_irr}
T_{\text{irr}}\equiv\int_0^{V_\text{irr}}dV'Q=\frac{C_{\text{tot}}}{2}\left(V_{\text{irr}}\right)^{2}\ .
\end{equation}

With this choice, the remaining electric field $\mathbf{E}_{\dot{B}}$
must be captured by the vector potential, which we can now uniquely
define as
\begin{equation}
\mathbf{A}_{\text{irr}}=-\mathbf{F}\left[\mathbf{B}\right],\label{eq:A_from_E}
\end{equation}
cf. Eq.~(\ref{eq:E_functional}) and subsequent discussion. When applied to the interior, Eq.~\eqref{eq:A_from_E} states that the vector potential in the interior corresponds to the London gauge, $\mathbf{A}=-\mu_0\lambda^2\mathbf{j}$, whereas in the exterior, it
has to satisfy $\boldsymbol{\nabla}\times\mathbf{A}_{\text{irr}}=\mathbf{B}$,
with the additional constraints $\boldsymbol{\nabla}\cdot\mathbf{A}_{\text{irr}}=0$
(Coulomb gauge), and 
\begin{equation}
\mathbf{n}_{l}\times\mathbf{A}_{\text{irr}}\biggr|_{\mathbf{x}\in\mathcal{S}_l}=-\lambda^2\mathbf{n}_l\times\left(\nabla\times \mathbf{B}\right)\biggr|_{\mathbf{x}\in\mathcal{S}_{l}}\ .\label{eq:A_condition}
\end{equation}
In addition, Eq.~\eqref{eq:A_from_E} implies that the integral of $\left.\mathbf{n}_i\cdot\mathbf{A}_\text{irr}\right|_{\mathbf{x}\in\mathcal{S}_i}$ over the island surface must vanish, in analogy to the surface charge discussion for $\mathbf{E}_{\dot{B}}$. All of these properties are obviously inherited from $\mathbf{E}_{\dot{B}}$, and
therefore render the solution for $\mathbf{A}_{\text{irr}}$ unique.
Crucially, Eq.~\eqref{eq:A_condition} implies that the tangential component of the vector potential in the exterior must equal the interior vector potential at the surface, when the latter is computed in the London gauge. The longitudinal part on the other hand is discontinuous, just like $\mathbf{E}_{\dot{B}}$. In essence, one might consider our result as a prescription to continue the London gauge to the exterior of the superconductor. 

Let us point out that in analogy to the discussion of $\mathbf{E}$, we may likewise simplify the computation of $\mathbf{A}$ by setting $\lambda\rightarrow0$, under the assumption that the relevant length scales of the device geometry are larger than the London penetration depth. This leads to the simplified condition
\begin{equation}
\mathbf{n}_{l}\times\mathbf{A}_{\text{irr}}\biggr|_{\mathbf{x}\in\mathcal{S}_l}=0,\label{eq:A_condition_simple}
\end{equation}
following directly from Eq.~\eqref{eq_E_boundary_simple}. Consequently, when the approximation $\lambda\rightarrow0$ is justified, the irrotational gauge loses its connection to the standard London gauge because the interior vector potential becomes irrelevant.

The phases that appear inside the potential energy in this irrotational
gauge at the Josephson junction $k$ are then given through the
line integrals
\begin{equation}
\phi_{k,\text{irr}}\left(t\right)=\frac{2\pi}{\Phi_{0}}\int_{\mathcal{L}_{Jk}}d\mathbf{l}\cdot\mathbf{A}_{\text{irr}}\ ,\label{eq:phi_irr}
\end{equation}
where we take $\mathcal{L}_{Jk}$ to be the shortest path across junction $k$. Note that since $\mathbf{A}_\text{irr}$ is by construction not curl-free, the resulting phase would in principle depend on the path, such that this choice needs to be justified. Large deviations from this shortest path are not considered because they are extremely unlikely, as can be seen, e.g., by means of standard path integral considerations. Small deviations from the shortest path (where the Cooper pairs still take quantum paths within the vicinity of the junction) leave the integral in Eq.~\eqref{eq:phi_irr} approximately invariant provided that the magnetic flux penetrating the junction is much smaller than the flux quantum. If fashioned for quantum hardware purposes, Josephson junctions are likely to satisfy this constraint (to avoid, e.g., Fraunhofer diffraction effects~\cite{tinkhambook}). 

For a SQUID with two Josephson junctions we thus find the Lagrangian
in the irrotational gauge, with $V_\text{irr}=2e\dot{\varphi}$,
\begin{multline}
L_{\text{irr}}=\frac{C_{\text{tot}}}{2}\left(\frac{\dot{\varphi}}{2e}\right)^{2}+E_{J1}\cos\left(\varphi+\phi_{1,\text{irr}}\right)\\\label{eq_L_irr}
+E_{J2}\cos\left(\varphi+\phi_{2,\text{irr}}\right)\ ,
\end{multline}
which represents the main result of this work. Indeed, from Eq.~\eqref{eq_L_irr}, we find the charge as the canonically conjugate momentum $Q=2en=2e\partial_{\dot{\varphi}}L_\text{irr}=C_\text{tot} \dot{\varphi}/2e$ as defined in Eq.~\eqref{eq_Q_vs_V_irr}. By means of a standard Legendre transformation, and the quantization of charge and phase, $\varphi,n\rightarrow\widehat{\varphi},\widehat{n}$ with $[\widehat{\varphi},\widehat{n}]=i$, we arrive at the sought-after Hamiltonian $\widehat{H}_\text{irr}(t)$ in the irrotational gauge:
\begin{multline}\label{eq_H_irr}
\widehat{H}_\text{irr}=\frac{\left(2e\right)^{2}}{2C_{\text{tot}}}\widehat{n}^{2}
-E_{J1}\cos\left(\widehat{\varphi}+\phi_{1,\text{irr}}\right)\\
-E_{J2}\cos\left(\widehat\varphi+\phi_{2,\text{irr}}\right)\ .
\end{multline}

Let us make three important remarks. First, here
we propose a procedure to obtain the correct Lagrangian which strongly
generalizes the work of Ref.~\citep{You2019}. Reference~\citep{You2019} invokes initially
separate degrees of freedom for each junction, taking separate capacitances
for each junction as the input. These degrees of freedom are then
reduced by means of appropriate constraints in the Lagrangian. While
this is a valid strategy under the hypothesis that the capacitances
of the individual junctions are known and well-defined, our approach
generalizes their effort beyond such simple geometries, and actually
provides a procedure which not only gives the correct Lagrangian
for specified junction capacitances, but in fact can be taken to provide also the correct
junction capacitances \textit{themselves}, by matching the solutions
for $\phi_{k,\text{irr}}\left(t\right)$ with the relations involving $C_{k}$,
see Eqs.~(\ref{eq:H_alpha}) and (\ref{eq:alpha_C}). In fact, this
general approach will lead to anomalous capacitances (negative or
even time-dependent) already for quite standard circuit and magnetic
field models, as we will discuss in the following.

Second, we would like to stress that even though the phases $\phi_{k,\text{irr}}$
are computed through local line integrals, they are still guaranteed
to contain full information of the flux enclosed by the two
Josephson junctions (which is a nonlocal property). This fact can be illustrated as follows. The path $\mathcal{L}_{Jk}$ across the junction, as it appears in Eq.~\eqref{eq:phi_irr} can first be continued in a straight line (normal to the surfaces), such that the starting and end points are well within the conductors, where $\mathbf{E}_{\dot{B}}$ and thus $\mathbf{A}_\text{irr}$ are zero. Since the interior $\mathbf{E}$- and $\mathbf{A}$-fields are purely tangential, this path modification does not change the integral in Eq.~\eqref{eq:phi_irr}. Then, these two paths can be joined by additional paths which reside entirely within the region of the conductors where $\mathbf{A}_\text{irr}=0$, to form a closed loop. These additional line integrals must therefore vanish, such that $\phi_{2,\text{irr}}-\phi_{1,\text{irr}}$
necessarily equates to $2\pi\Phi/\Phi_{0}$ with $\Phi$ defined in Eq. (\ref{eq:flux}). 

Third, we note that the procedure is readily generalizable to larger
circuits that include many additional circuit elements. For instance, many islands carrying island charges $Q_i$ can be included by expressing the electric field $\mathbf{E}_Q$ in Eq.~\eqref{eq_E_field_scaling} as a linear superposition, $\mathbf{E}_Q=\sum_i Q_i\mathbf{e}_i$. With this extension, the formulation of the irrotational gauge for the scalar and vector
potentials remains the same; in particular, for the latter, that would
be the Coulomb gauge plus the boundary condition, Eq.~\eqref{eq:A_condition}. For an arbitrary number
of Josephson junctions, each junction $k$ adds a contribution $-E_{Jk}\cos\left(\varphi_{k}+\phi_{k,\text{irr}}\right)$
to the potential energy, with $\dot{\varphi}_{k}=2eV_{k}$ where $V_{k}$
is the voltage difference between the islands connected by junction
$k$, and $\phi_{k,\text{irr}}$ still defined as in Eq.~(\ref{eq:phi_irr}).

We note one nontrivial extension: in general, there could be an additional
capacitively coupled gate with a voltage $V_{g}$, inducing offset gate charges
$n_{g}=C_{g}V_{g}$, $\widehat{n}\rightarrow\widehat{n}+n_{g}$. In
this case, the Hamiltonian would of course no longer be purely quadratic in $\widehat{n}$.
This however necessitates only a minor extension of the above prescription
for finding the irrotational gauge for $\mathbf{A}$. Namely, any
gate gives rise to an additional source term in the electromagnetic
problem, $\mathbf{E}=\mathbf{E}_{Q}+\mathbf{E}_{\dot{B}}+\mathbf{E}_{V_{g}}$, where the new field $\mathbf{E}_{V_{g}}$ is linearly independent of the others. Hence, in order
to determine $\mathbf{A}_{\text{irr}}$ we simply have to set all
gate voltages to zero, $V_{g}=0$ (that is, all gate voltages equal
to a common ground), resulting again in $\mathbf{E}_{\dot{B}}=-\mathbf{F}\left[\dot{\mathbf{B}}\right]$,
allowing us to proceed as above.

Finally, let us point out that there is a straightforward generalization of our gauge constraints, which eventually allow for terms linear in $\widehat{n}$ in the Hamiltonian. Namely, the physical charge $Q$ in Eq.~\eqref{eq_island_charge} may be supplemented with an auxiliary (and in general time-dependent) shift $Q\rightarrow Q-Q_0(t)$, leading to a new solution $\mathbf{E}_Q'=\left(Q-Q_0\right)\mathbf{e}$. Since this shifted charge is not physical (unlike an offset charge induced by a physical gate, as discussed above), it has to be subtracted again as an extra term in a new $\mathbf{E}_{\dot{B}}'$, such that $\mathbf{E}_Q'+\mathbf{E}_{\dot{B}}'=\mathbf{E}_Q+\mathbf{E}_{\dot{B}}$. Continuing the subsequent steps of the above recipe with these new fields will give rise to a new Hamiltonian $\widehat{H}_\text{irr}'=\widehat{U}\widehat{H}_\text{irr}\widehat{U}^\dagger-i\widehat{U}\partial_t \widehat{U}^\dagger$, where the time-dependent unitary $\widehat{U}$ depends on the integral of the shifted charge, $\int^t dt'Q_0$. For details, see Appendix~\ref{app_T_and_general_gauge}. At any rate, such a \textit{variably-shifted irrotational gauge} connects a more general choice of decomposing the electric field (and thus finding a different gauge for the resulting scalar and vector potentials) with a more general class of Hamiltonians describing the same driven circuit, see also Eqs.~(13-14) of \cite{You2019}.

\subsection{Neglecting junction filaments\label{subsec:junction-filaments}}

\begin{figure}
\centering

\includegraphics[width=0.8\columnwidth]{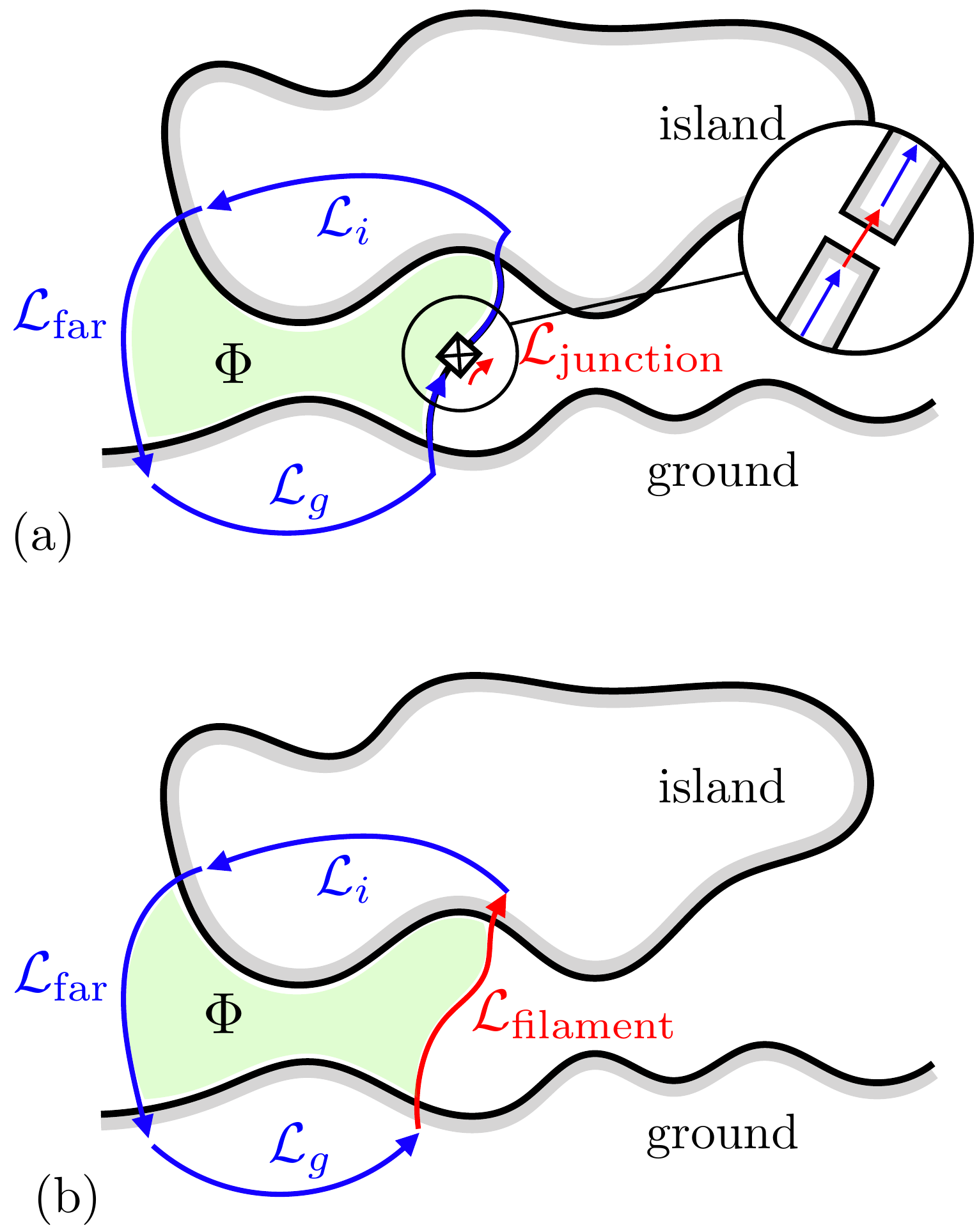}

\caption{Illustration of the argument why junction filaments can be neglected.
In (a), the circuit is drawn including the actual junction, whereas
in (b) the junction is removed in the model. Instead, there is a path
$\mathcal{L}_{\text{filament}}$ (in red), which takes a path equivalent
to the filament. \label{fig:filament_argument}}

\end{figure}

While the above procedure is general, we here discuss an important
simplification, which we expect to be instrumental for reducing
computational effort. As mentioned in the introduction, Josephson
junctions connecting individual bulk conductors often have a filamentary structure
[as indicated in Figs.~\ref{fig_intro}(a) and \ref{fig:filament_argument}(a)], because the Josephson junctions are fabricated using the Niemeyer-Dolan technique~\citep{Niemeyer_1976,Dolan_1977}. We here argue under which conditions
the filaments can be neglected altogether in obtaining field solutions.

The argument requires two main assumptions. (i) The filaments' contribution
to the capacitance is negligible. Thus, their presence will distort
the solutions of the electric fields $\mathbf{E}_{\dot{B}}({\bf r},t)$ and $\mathbf{E}_{Q}(\bf{r})$
only locally, whereas far away from the junctions, these fields will be the
same as if the filaments were not present. We can therefore always
expect to find a path $\mathcal{L}_{\text{far}}$ connecting ground
and island, where the field solutions are the same as when the
junction filaments are absent, see Figs.~\ref{fig:filament_argument}(a) and (b),
respectively. (ii) While it may be expected, that the filaments have
their own Meissner effect (i.e., we assume the London penetration
depth to be smaller than the filament dimensions), due to the filaments
being small, their presence does not lead to a significant distortion
of $\mathbf{B}({\bf r},t)$.

Equipped with these two assumptions, let us consider a closed path
for the circuit with the filament (Fig.~\ref{fig:filament_argument}(a))
and without it (Fig.~\ref{fig:filament_argument}(b)). As depicted in the figures [and similar to the discussion below Eq.~\eqref{eq_H_irr}], the paths $\mathcal{L}_\text{far}$, $\mathcal{L}_\text{junction}$, and $\mathcal{L}_\text{filament}$ are chosen such that they pierce the conductors (normal to the surface) up to a suffiently large length, such that the Meissner screening currents decay (gray area).
In general, the
two circuits will have different field solutions, $\mathbf{E}_{\dot{B}}$
and $\widetilde{\mathbf{E}}_{\dot{B}}$ as well as $\mathbf{A}_{\text{irr}}$
and $\widetilde{\mathbf{A}}_{\text{irr}}$ (as per the irrotational gauge defined
in Sec. \ref{irrotational_gauge}). Due to (ii), as long as $\mathcal{L}_{\text{filament}}$
in Fig. \ref{fig:filament_argument}(b) is chosen, such that the closed
path covers the exact same area as the one in Fig. \ref{fig:filament_argument}(a),
the enclosed flux $\Phi$ will be the same. Hence,
\begin{equation}
\oint_{\mathcal{L}_{\text{far}}+\mathcal{L}_{i}+\mathcal{L}_{g}+\mathcal{L}_{\text{junction}}}\!\!\!\!\!\!\!\!\!\!\!\!\!\!\!d\mathbf{l}\cdot\mathbf{A}_{\text{irr}}=
\oint_{\mathcal{L}_{\text{far}}+\mathcal{L}_{i}+\mathcal{L}_{g}+\mathcal{L}_{\text{filament}}}\!\!\!\!\!\!\!\!\!\!\!\!\!\!\!d\mathbf{l}\cdot\widetilde{\mathbf{A}}_{\text{irr}}.
\end{equation}
Because the vector potential in the irrotational gauge is zero
inside the conductors, the contributions
of these paths, $\mathcal{L}_{i,g}$, have
to disappear, leaving us with
\begin{equation}
\begin{split}
\int_{\mathcal{L}_{\text{far}}}d\mathbf{l}\cdot\mathbf{A}_{\text{irr}}+\int_{\mathcal{L}_{\text{junction}}}d\mathbf{l}\cdot\mathbf{A}_{\text{irr}}=\\ \int_{\mathcal{L}_{\text{far}}}d\mathbf{l}\cdot\widetilde{\mathbf{A}}_{\text{irr}}+\int_{\mathcal{L}_{\text{filament}}}d\mathbf{l}\cdot\widetilde{\mathbf{A}}_{\text{irr}}.
\end{split}
\end{equation}
Finally, in accordance with assumption (i), sufficiently far from
the junction the vector field solution is the same for both circuits,
such that the first term on either side of the equation cancels, resulting
in
\begin{equation}\label{eq_A_filament}
\int_{\mathcal{L}_{\text{junction}}}d\mathbf{l}\cdot\mathbf{A}_{\text{irr}}=\int_{\mathcal{L}_{\text{filament}}}d\mathbf{l}\cdot\widetilde{\mathbf{A}}_{\text{irr}},
\end{equation}
which thus shows that one may neglect the actual junction filaments
for the computation of the field solutions, as long as an equivalent
path, $\mathcal{L}_{\text{filament}}$, is chosen. We note that the
above proof can be extended to an arbitrary number of junctions, as
long as (i) and (ii) remain valid.
The observation described in Eq.~\eqref{eq_A_filament} can in some sense be interpreted as a ``lightning rod effect'' of the filament. Namely, if it holds, it means that the vector potential at the junction is increased by a factor $L_\text{filament}/d\gg 1$ compared to the solution in the absence of the filament (where $d$ is the junction thickness and $L_\text{filament}$ is the length of the filament path $\mathcal{L}_\text{filament}$).

\section{Anomalous capacitances\label{sec:anomalous-capacitances}}

We now apply the above general scheme to a concrete SQUID geometry that allows for simple analytic solutions of the Maxwell equations.
We will show that for this concrete case, a mapping from the general
circuit to a lumped element model, see Fig.~\ref{fig_intro}, cannot
be accomplished unless negative or even time-dependent capacitances
are assigned to the individual junctions.

\begin{figure}
\includegraphics[width=1\columnwidth]{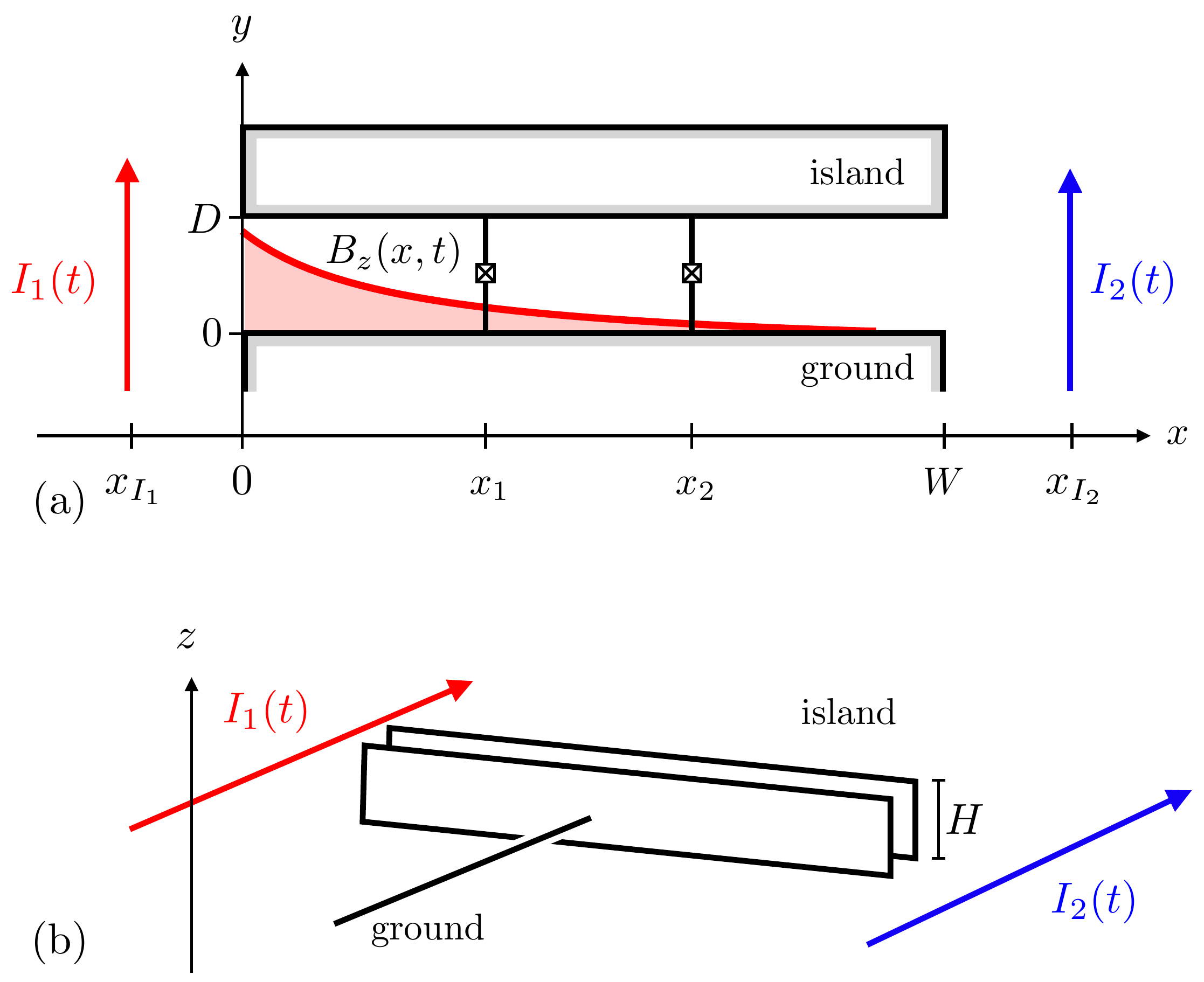}

\caption{Sketch of the simple SQUID circuit with two magnetic field sources. The setup as shown in this figure leads to all the predicted effects in the main text, such as negative capacitances, or time-dependent capacitances. (a) Top view of the SQUID circuit. The ground and island form a parallel
plate capacitor with distance $D$ ($y$-direction) and width $W\gg D$
($x$-direction). Two current wire sources $I_{1}$
(red) and $I_{2}$ (blue) produce magnetic fields. We indicate the spatial distribution of the field for the first source (red curve). The Josephson
junctions are placed at positions $x_{1}$ and $x_{2}$, $0<x_{1,2}<W$
while the sources are placed at $x_{I_{1}}<0$ and $x_{I_{2}}>W$. Note that for illustration purposes, the $y$-axis is exaggerated; in the actual model both the plate distance $D$ and the plate thicknesses should be much smaller than any of the length scales in $x$ direction. (b) In $z$-direction the planes are of a height
$H$ which is large with respect to the distance, $H\gg D$, but small
with respect to the width $H\ll W$.
 \label{fig:1D_circuit}}
\end{figure}

\subsection{Explicit solution for irrotational vector potential\label{subsec:explicit_solution_1D}}

We assume that the ground and island form a parallel plate capacitor
with width $W$ and separation $D$ ($x$- and $y$-directions, respectively),
see Fig.~\ref{fig:1D_circuit}. In $z$-direction the plates extend
to height $H$. Both $H,W\gg D$, such that fringe effects can be
neglected. The total capacitance is therefore the standard expression
for the parallel plate capacitor $C_{\text{tot}}=\epsilon HW/D$. 

In order to derive the Hamiltonian, we need to know the vector potential
$\mathbf{A}_{\text{irr}}$ within the thin volume separating the two plates.
As for the magnetic field, we assume for simplicity that it is created
by two wires carrying a time-dependent current $I_{1,2}\left(t\right)$,
oriented in $y$-direction, at positions $x=x_{I_{1,2}}$ and $z=z_I$, which we set to be $z_I=H/2$ (see Fig.~\ref{fig:1D_circuit}). It is therefore of the form
\begin{equation}\label{eq_B_field}
\mathbf{B}\approx\left(\begin{array}{c}
B_{x}\left(x,z,t\right)\\
0\\
B_{z}\left(x,z,t\right)
\end{array}\right),
\end{equation}
with the components given as per Amp\`{e}re's law,
\begin{align}
B_{x}\left(x,z,t\right)&=\frac{\mu I_{1}\left(t\right)}{2\pi r_1^2}\left(z-z_I\right)+\frac{\mu I_{2}\left(t\right)}{2\pi r_2^2}\left(z-z_I\right)\\
B_{z}\left(x,z,t\right)&=-\frac{\mu I_{1}\left(t\right)}{2\pi r_1^2}\left(x-x_{I_1}\right)-\frac{\mu I_{2}\left(t\right)}{2\pi r_2^2}\left(x-x_{I_2}\right),\label{eq:bz_I1_I2}
\end{align}
with the distance from the current source $I_{1,2}$, $r_{1,2}^2=\left(x-x_{I_{1,2}}\right)^2+\left(z-z_I\right)^2$.

The solution for the electric field, $\mathbf{E}_{\dot{B}}$, fulfilling
the conditions detailed in Sec. \ref{unique_fields}, is given as follows (see Appendix~\ref{app:parallel_plate_derivations} for the derivation). In the interior of the conductor plates, the magnetic field decays exponentially on the length scale $\lambda$. The Meissner screening currents give rise to an ac electric field via the first London equation, $\mathbf{E}=\mu\lambda^{2}\partial_{t}\mathbf{j}$, which can be given for the two capacitor plates as
\begin{equation}\label{eq_E_upper_plate}
    \mathbf{E}_{\dot{B}}\left(y>D\right)=-\lambda\left(\mathbf{n}_{y}\times\dot{\mathbf{B}}\right)e^{-\frac{y-D}{\lambda}}\ .
\end{equation}
respectively
\begin{equation}\label{eq_E_lower_plate}
    \mathbf{E}_{\dot{B}}\left(y<0\right)=\lambda\left(\mathbf{n}_{y}\times\dot{\mathbf{B}}\right)e^{\frac{y}{\lambda}}\ ,
\end{equation}
where $\mathbf{n}_y=(0,1,0)^T$. The condition $\nabla\cdot\mathbf{E}_{\dot{B}}=0$ for the interior is a consequence of the likewise source-free screening currrent $\nabla\cdot\mathbf{j}=0$, which in turn follows from $\nabla\times\mathbf{B}=0$ for the magnetic field in the exterior (away from the wires).

The above solutions for $\mathbf{E}_{\dot{B}}$, evaluated at the plate surfaces $y=0$ and $y=D$, respectively, provide the right-hand side of the boundary conditions in Eq.~\eqref{eq_E_boundary}, such that we can now solve for the volume in between the plates, $0<y<D$ and $0<z<H$. For this particular geometry, it turns out to be helpful to explicitly decompose the field into a longitudinal (oriented along $y$) and tangential (parallel to the plate surfaces) component, $\mathbf{E}_{\dot{B}}(0<y<D)=\mathbf{E}_L+\mathbf{E}_T$. We find for the longitudinal field $\mathbf{E}_L=E_{L,y}\mathbf{n}_y$ with,
\begin{align}\nonumber
    E_{L,y}&=\left(1+\frac{2\lambda}{D}\right)\int_{0}^{W}\frac{dx''}{W}\int_{0}^{H}\frac{dz''}{H}\int_{x}^{x''}dx'b_{z}\left(x',z,z''\right)\\ \label{eq_E_longitudinal}&+\left(1+\frac{2\lambda}{D}\right)\int_{0}^{H}\frac{dz''}{H}\int_{0}^{W}\frac{dx''}{W}\int_{z''}^{z}dz'b_{x}\left(x,x'',z'\right)\ ,
\end{align}
where $2b_{x}\left(x,x'',z'\right)=B_{x}\left(x,z'\right)+B_{x}\left(x'',z'\right)$ as well as $2b_{z}\left(x',z,z''\right)=B_{z}\left(x',z\right)+B_{z}\left(x',z''\right)$. The tangential field reads
\begin{equation}\label{eq_tangential_E}
    \mathbf{E}_{T}=\lambda\left(\mathbf{n}_{y}\times\dot{\mathbf{B}}_{\text{ext}}\right)\frac{D-2y}{D}\ .
\end{equation}
Here, the tangential field continuously matches the solution for the interior of the conductors, given in Eqs.~\eqref{eq_E_upper_plate} and~\eqref{eq_E_lower_plate}. The longitudinal field abruptly goes to zero at $y=0,D$, thus giving rise to a surface charge, as defined in Eq.~\eqref{eq_surface_charge}, induced by $\dot{\mathbf{B}}\neq 0$. According to Sec.~\ref{irrotational_gauge}, this surface charge must integrate to zero. This can be verified when integrating $\mathbf{E}_L$ over $x$ and $z$. In the first line, there will be the total integral $\int_{0}^{W}dx\int_{0}^{W}dx''$ applied to a function of the form $\int_{x}^{x''}dx'\ldots$, which is obviously antisymmetric upon exchanging $x\leftrightarrow x''$. The argument applies similarly for $z$ in the second line.
Importantly, in accordance with Sec. \ref{subsec:junction-filaments},
we here neglect the influence of the junction filaments for the
solution of $\mathbf{E}_{\dot{B}}$.

Following the lines of Sec.~\ref{irrotational_gauge}, the vector potential is obtained in a straightforward fashion by taking the solutions for the $\mathbf{E}$-field in Eqs.~(\ref{eq_E_upper_plate}-\ref{eq_tangential_E}), and replacing $\dot{\mathbf{B}}\rightarrow-\mathbf{B}$. We are now able to make some crucial simplifications. First of all, we notice that if the capacitor plate separation $D\gg \lambda$, we may simplify $\lambda/D\approx 0$ in Eq. (\ref{eq_E_longitudinal}), and $\mathbf{E}_T\approx 0$. The remaining terms of the longitudinal field then correspond to the field solutions with the simplified boundary condition, Eq.~\eqref{eq_E_boundary_simple}. We thus explicitly show at the example of the parallel plate capacitor, that $\lambda\rightarrow 0$ is justified if the relevant length scales of the device geometry exceed $\lambda$, as foreshadowed in Sec.~\ref{irrotational_gauge}.

In addition, we assume a quasi one-dimensional parallel plate capacitor, $W\gg H\gg D$, such that we may simplify $B_x\approx 0$ and $B_z(x,z)\approx B_z(x)$, that is, the magnetic field is mainly oriented in $z$-direction, and depends only on $x$~\footnote{For this approximation to remain valid, we always need to consider distances between the wires and the capacitor larger than $H$. Hence, whenever we refer to the wires being close to the capacitor, we always mean distances smaller than $W$, but larger than $H$.}. As a consequence, $b_x\approx 0$ and $b_z(x',z,z'')\approx B_z(x)$. For the resulting
irrotational gauge, as defined in Sec. \ref{irrotational_gauge}, we find,
\begin{equation}
\mathbf{A}_{\text{irr}}=-\int_{0}^{W}\frac{dx'}{W}\int_{x}^{x'}dx''B_{z}\left(x'',t\right)\mathbf{n}_y.\label{eq:A_irr_1D}
\end{equation}
The phases at the two Josephson junctions return as a result,
\begin{equation}
\phi_{k,\text{irr}}(t)=-\frac{2\pi D}{\Phi_{0}}\int_{0}^{W}\frac{dx'}{W}\int_{x_{k}}^{x'}dx''B_{z}\left(x'',t\right).\label{eq:phi_irr_1D}
\end{equation}
Based on this solution, we now may examine the
extent to which a mapping onto a lumped element circuit Fig.~\ref{fig_intro}(b)
can be achieved. Given the discussion in Sec. \ref{the_state_of_the_art},
the mapping must give rise to junctions with the effective capacitances [see Eqs.~(\ref{eq:H_alpha},\ref{eq:alpha_C})]
\begin{align}
C_{\text{eff},1}\left(t\right) & =\frac{\phi_{2,\text{irr}}\left(t\right)}{\phi_{2,\text{irr}}\left(t\right)-\phi_{1,\text{irr}}\left(t\right)}C_{\text{tot}},\label{eq:C1_eff}\\
C_{\text{eff},2}\left(t\right) & =-\frac{\phi_{1,\text{irr}}\left(t\right)}{\phi_{2,\text{irr}}\left(t\right)-\phi_{1,\text{irr}}\left(t\right)}C_{\text{tot}}.\label{eq:C2_eff}
\end{align}
In what follows, we will derive explicit expressions for $C_{\text{eff},k}$ starting from above Eqs.~\eqref{eq:C1_eff} and~\eqref{eq:C2_eff}. As we will see, the spatial inhomogeneity of the magnetic
fields is at the heart of the nontrivial behaviour for 
$C_{\text{eff},k}$. We will argue, in particular, that for a single
source, negative effective capacitances $C_{\text{eff},k}$ emerge.
Two independent sources lead to time-dependent capacitances.

\subsection{Negative junction capacitances and qubit relaxation rates}

Let us first consider the case of only one source, $I\equiv I_{1}$
and set $I_{2}=0$. In this case, the vector potential in the irrotational
gauge is, according to Eq. (\ref{eq:A_irr_1D}), in the limit $\left|x_{I_{1}}\right|\ll W$ (while still $|x_{I_1}|\gg H$)
\begin{equation}
\mathbf{A}_{\text{irr}}\approx-\frac{\mu I\left(t\right)}{2\pi}\left[\ln\left(\frac{W}{x}\right)-1\right]\mathbf{n}_y.
\end{equation}
The resulting phases $\phi_{\text{irr},k}$ depend on time only through
a global prefactor, $\phi_{\text{irr},k}\sim I\left(t\right)$ which
cancels in Eqs. (\ref{eq:C1_eff}) and (\ref{eq:C2_eff}). The resulting
capacitances are therefore time-independent,
\begin{align}
C_{\text{eff},1} & \approx\frac{\ln\left(\frac{W}{x_{2}}\right)-1}{\ln\left(\frac{x_{1}}{x_{2}}\right)}C_{\text{tot}}\\
C_{\text{eff},2} & \approx-\frac{\ln\left(\frac{W}{x_{1}}\right)-1}{\ln\left(\frac{x_{1}}{x_{2}}\right)}C_{\text{tot}}.
\end{align}
However, they can be negative. Notably, already for a completely
symmetric junction placement $x_{1,2}=W/2\mp\delta x$, we find $C_{\text{eff},2}<0$
and $C_{\text{eff},1}>C_{\text{tot}}$ if
\begin{equation}
\frac{\delta x}{W}<\frac{1}{2}-\frac{1}{e}\approx0.13,
\end{equation}
where $e$ here denotes Euler's number (not to be confused with the
elementary charge). Note that this corresponds to a very strong asymmetry
in the junction capacitances, even though the circuit geometry itself
is perfectly symmetric. The effect stems solely from the spatial asymmetry
of the time-dependent magnetic field.

Crucially, the presence of negative capacitances strongly affects
the accurate prediction of the transition rates between the ground
and excited states of the Hamiltonian. Along the same lines as in
Ref.~\citep{You2019}, let us consider fluctuations of the flux $\phi=\phi_{0}+\delta{\phi}$
around an equilibrium value $\phi_{0}$ (which could in this concrete
case correspond to current fluctuations). These give rise to the relaxation
rate (for simplicity, for symmetric junctions, $E_{J1}=E_{J2}=E_{J}$)
\begin{equation}
\begin{split}
\Gamma_{m\rightarrow m'}\sim\left(\frac{C_{1,\text{eff}}-C_{2,\text{eff}}}{C_{\text{tot}}}\right)^{2}S_{\delta\phi}\left(\omega_{mm'}\right)\\ \times E_{J}^{2}\left|\left\langle m\right|\sin\left(\varphi\right)\left|m'\right\rangle \right|^{2},
\end{split}
\end{equation}
where $m$ is the index that counts the eigenstates of $H$, and the
phase noise power spectrum $S_{\delta\phi}\left(\omega\right)$ is
taken at the frequency corresponding to the transition energy $\omega_{mm'}=\epsilon_{m}-\epsilon_{m'}$.
We see now that ignoring the possibility of anomalous capacitances
leads to a potential massive underestimation of the transition rates.
While for regular capacitances, we find 
\begin{equation}
0<\left(\frac{C_{1,\text{eff}}-C_{2,\text{eff}}}{C_{\text{tot}}}\right)^{2}<1,
\end{equation}
for anomalous capacitances, it may happen that
\begin{equation}
\left(\frac{C_{1,\text{eff}}-C_{2,\text{eff}}}{C_{\text{tot}}}\right)^{2}\gg1.
\end{equation}
Note though, that negative capacitances do \textit{not} lead to a
breakdown of the perturbation theory. For concreteness,
consider the above model of a single wire source, in the
limit where the separation of the two junctions is small, $\delta x\ll W$.
Then,
\begin{equation}
\Gamma_{m\rightarrow m'}\sim\left(\frac{W}{\delta x}\right)^{2}S_{\delta\phi}\left(\omega_{mm'}\right)E_{J}^{2}\left|\left\langle m\right|\sin\left(\varphi\right)\left|m'\right\rangle \right|^{2}.
\end{equation}
On the one hand, the prefactor $\left(W/\delta x\right)^{2}$ (due to negative capacitances) is large. On the other hand, for $\delta x\rightarrow0$ (where the two junctions
essentially merge to one), the total flux enclosed by the two junctions approaches zero, too, $\delta\phi\rightarrow0$, linearly with $\delta x$. Hence, the phase noise power spectrum $S_{\delta\phi}$
goes to zero at the same rate the prefactor $\left(W/\delta x\right)^{2}$
diverges, such that the product $\left(W/\delta x\right)^{2}S_{\delta\phi}$
remains finite.

To summarize, without appropriately taking into account the spatial
distribution of the magnetic field, one might have naively expected
that with the total phase $\phi$, respectively, its fluctuations
$\delta\phi$ going to zero, the circuit may loose its sensitivity
to the noise emitted by the magnetic field source. We here show that
this is not so; depending on the spatial details of the magnetic field
the qubit relaxation rate due to magnetic noise remains relevant even
if the area enclosed by the SQUID is small.

\subsection{Geometric phase generated by time-dependent capacitances}

Now we consider a driving of the device by means of two wires, as
shown in Fig.~\ref{fig:1D_circuit}, such that in the expression for $B_{z}$ in Eq.~(\ref{eq:bz_I1_I2}),
we keep both $I_{1}$ and $I_{2}$ nonzero. For simplicity, we consider again a fully symmetric
geometry for both the circuit, $x_{1,2}=W/2\mp\delta x$, and the
current sources approach the edges of the capacitor from both sides, $x_{I_{1}}\rightarrow0$
and $x_{I_{2}}\rightarrow W$ (again, provided that $|x_{I_1}|,|W-x_{I_2}|\gg H$). We find for the vector potential
\begin{multline}
\mathbf{A}_{\text{irr}}\approx-\frac{\mu I_{1}\left(t\right)}{2\pi}\left[\ln\left(\frac{W}{x}\right)-1\right]\mathbf{n}_y\\
-\frac{\mu I_{2}\left(t\right)}{2\pi}\left[\ln\left(\frac{W}{W-x}\right)-1\right]\mathbf{n}_y.\label{eq:A_irr_1D_two_wires}
\end{multline}
For the resulting phases, we compactify the expressions by introducing
the decomposition  $\phi_{1,2}=\tfrac{1}{2}(\overline{\phi}\mp\phi)$, where
\begin{align}
\overline{\phi} & =2\frac{D\mu}{\Phi_{0}}\left[I_{1}\left(t\right)+I_{2}\left(t\right)\right]\left[1+\ln\left(\frac{1}{2}-\frac{\delta x}{W}\right)\right]\\
\phi & =-\frac{D\mu}{\Phi_{0}}\left[I_{1}\left(t\right)-I_{2}\left(t\right)\right]\ln\left(\frac{\frac{1}{2}-\frac{\delta x}{W}}{\frac{1}{2}+\frac{\delta x}{W}}\right).
\end{align}
When driving $I_{1}$ and $I_{2}$ independently, the Hamiltonian
is driven by two genuinely independent parameters, $\phi_{1}$ and
$\phi_{2}$, instead of just the single total phase enclosed by the
SQUID, $\phi_{2}-\phi_{1}$, as would be the case with a naive lumped-element
approach. In fact, as predicted, the result here can only be mapped
to a lumped element circuit of the form in Fig.~\ref{fig_intro}(b)
if the effective junction capacitances are allowed to be time dependent,
$C_{\text{eff},1,2}\left(t\right)=\frac{1}{2}\left(1\pm\overline{\phi}/\phi\right)C_{\text{tot}}$,
since the time-dependent prefactor for $\overline{\phi}/\phi\sim\left[I_{1}\left(t\right)+I_{2}\left(t\right)\right]/\left[I_{1}\left(t\right)-I_{2}\left(t\right)\right]$
does not cancel.

What is more, for independent currents $I_{1}$ and $I_{2}$ it can
very easily occur that $I_{1}-I_{2}=0$ while $I_{1}+I_{2}\neq0$
at certain moments in time, leading to singular capacitances. We
stress though, that this singularity is not a sign of a failure of
the theory: all system parameters in the Hamiltonian stay regular.
Instead, such singularities merely represent the failure to capture
the dynamics of the realistic system by means of the lumped element
approach in Fig. \ref{fig_intro}(b), when trying to decompose the total
capacitance into partial capacitances for each junction.

The presence of two explicit time-dependent parameters can be experimentally
accessed as follows. Consider a periodic ac driving
of the two currents. The two parameters $\phi_{1},\phi_{2}$,
or equivalently $\phi,\overline{\phi}$, enclose a finite area in
the 2D parameter space. In the adiabatic driving regime (when the
ac frequency is low with respect to the qubit frequency) a nontrivial
Berry phase may emerge as a consequence. When preparing the system
in a certain eigenstate $\left|m\right\rangle $, this Berry phase
may be expressed as
\begin{equation}
\mathcal{B}_{m}=2\,\text{Im}\iint d\phi d\overline{\phi}\,\partial_{\phi}\left\langle m\right|\partial_{\overline{\phi}}\left|m\right\rangle .
\end{equation}
In order to simplify further, let us focus on $E_{J1}=E_{J2}\equiv E_{J}$.
As will become clear in a moment, we should include
a stationary gate voltage in our system inducing a charge $n_{g}$
(see end of Sec.~\ref{irrotational_gauge}),
such that the Hamiltonian reads
\begin{equation}
\widehat{H}\left(t\right)=\frac{1}{2}E_{C}\left(\widehat{n}+n_{g}\right)^{2}-E_{J}\left[\phi\left(t\right)\right]\cos\left[\widehat{\varphi}-\frac{\overline{\phi}\left(t\right)}{2}\right],\label{eq:H_simple_case}
\end{equation}
with $E_{C}=\left(2e\right)^{2}/C_{\text{tot}}$ and $E_{J}\left(\phi\right)=2\cos\left(\phi/2\right)E_{J}$.
For further evaluation, it is helpful to reexpress the Berry curvature
as
\begin{equation}
\partial_{\phi}\left\langle m\right|\partial_{\overline{\phi}}\left|m\right\rangle =-\sum_{m'\neq m}\frac{\left\langle m\right|\partial_{\phi}\widehat{H}\left|m'\right\rangle \left\langle m'\right|\partial_{\overline{\phi}}\widehat{H}\left|m\right\rangle }{\left(\epsilon_{m}-\epsilon_{m'}\right)^{2}},
\end{equation}
where $\partial_{\phi}\widehat{H}\sim\cos\left(\widehat{\varphi}-\overline{\phi}/2\right)$
and $\partial_{\overline{\phi}}\widehat{H}\sim\sin\left(\widehat{\varphi}-\overline{\phi}/2\right)$.
Now, the importance of a finite gate voltage shift becomes obvious.
Observe that if $n_{g}=0$, the wave functions can be separated into two
subsets, one containing all wave functions which are symmetric, respectively
antisymmetric, in $\varphi$-space (with respect to $\overline{\phi}/2$),
resulting in a vanishing $\mathcal{B}_{m}$. But for finite (non-integer)
$n_{g}$ we can expect a finite Berry phase.

In fact, the Berry phase should have its largest value close to
$\phi\approx\pm\pi$ (up to multiples of $2\pi$), where the interference
of Josephson tunnelings across the two junctions is destructive, $E_{J}\cos\left(\phi/2\right)\ll E_{C}$,
while at the same time, keeping $n_{g}$ close to a charge degeneracy
point, $n_{g}\approx1/2$ (up to integer multiples), making sure that
Cooper pair transport is not fully suppressed. This corresponds to
the Cooper pair box regime \citep{Cottet_2002}, where it is easy
to find a good analytic approximation for the Berry curvature of the
even parity ground state $m=0$ (see Appendix \ref{app:Berry_curvature}),
\begin{equation}
2\text{Im}\partial_{\phi}\left\langle 0\right|\partial_{\overline{\phi}}\left|0\right\rangle =-\frac{1}{4}\partial_{\phi}\left(\frac{E_{C}\delta n_{g}}{\sqrt{E_{C}^{2}\delta n_{g}^{2}+4E_{J}^{2}\cos^{2}\left(\phi/2\right)}}\right),\label{eq:Berry_curvature}
\end{equation}
where $\delta n_{g}=n_{g}-1/2$ is the distance of $n_{g}$ away from the
charge degeneracy point. This result represents a measurable effect
of the highly nontrivial interaction of spatially asymmetric time-dependent
magnetic fields with a SQUID. Berry curvatures have been successfully
measured in superconducting qubits, see Ref. \citep{Abdumalikov_2013,Roushan_2014},
which is why we expect this effect to be readily observable.

\section{Refined lumped-element approach\label{sec:refined-approach}}

We have shown above that the ``naive'' lumped element approach
from Fig.~\ref{fig_intro} does not succeed in predicting the correct
system dynamics of a realistic SQUID geometry, unless one allows
for the possibility of anomalous (negative or time-dependent) capacitances. 

In this final section, we show that a description of the circuit with
regular, constant capacitances may still work to describe realistic
geometries, provided that the circuit is greatly refined. That is,
one needs to introduce a finely-meshed network of lumped elements
which can capture the spatial details of the externally applied field. In addition, capacitances need to be introduced at \textit{all} nodes, even (or especially) the ones which are not connected via Josephson junctions. 
By means of such a network, we show, for the above example of a 1D
SQUID, that our irrotational gauge procedure is equivalent to the
one developed by You et al. \citep{You2019}, when going to the continuum
limit. In addition, through this procedure we also include the internal
dynamics of the island (by describing it as a transmission line),
and thus provide an upper bound for the driving frequency, below which
the description of the island by means of a single independent degree of
freedom, $\left(\varphi,\dot{\varphi}\right)$, is justified.

\begin{figure}
\centering\includegraphics[width=1\columnwidth]{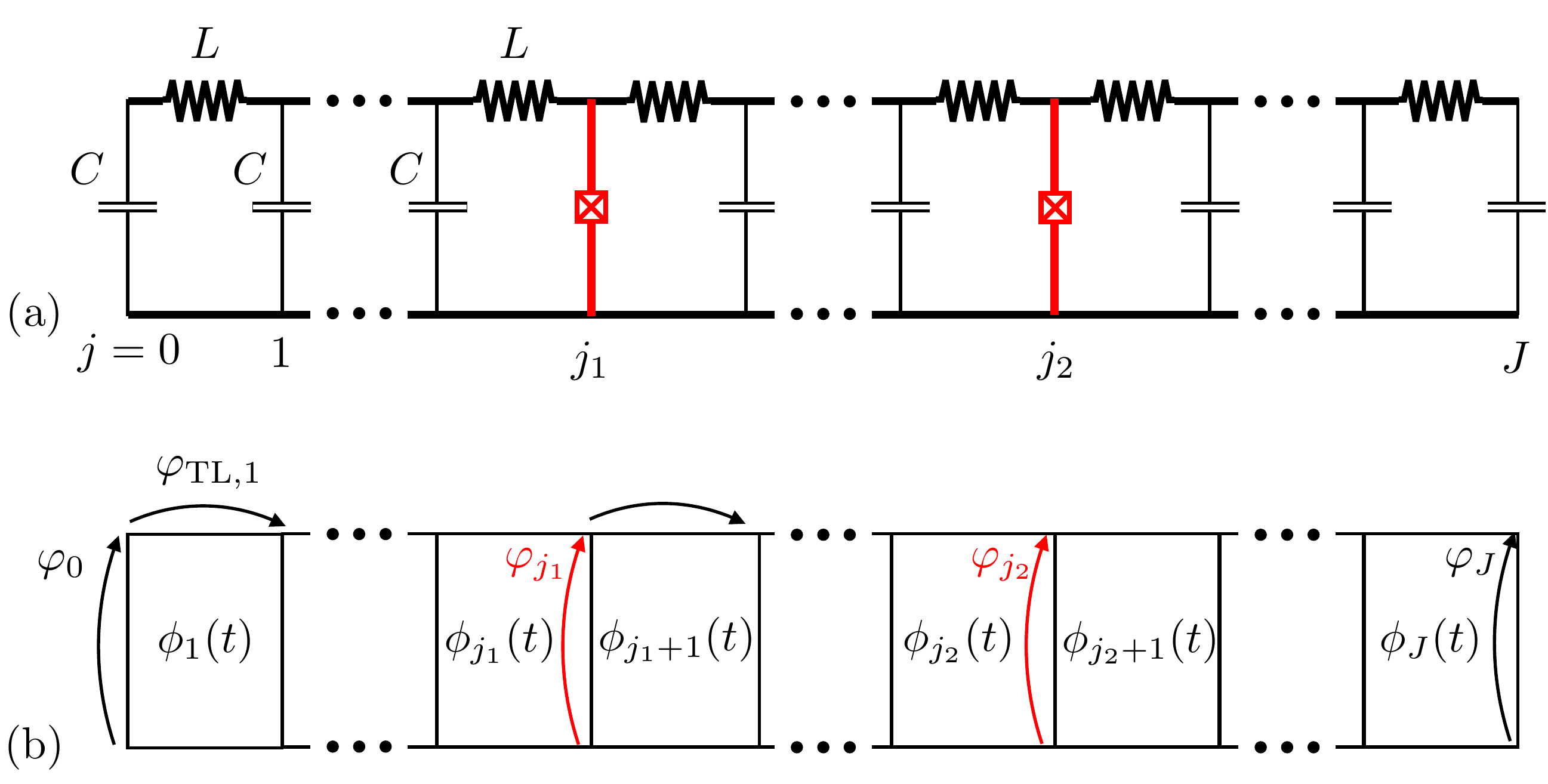}

\caption{(a) Finite element model to describe the SQUID geometry in Fig.~\ref{fig:1D_circuit}(a).
The upper arm of the SQUID is described through a transmission line,
whereas the lower arm is simply a ground. The Josephson junctions
(red) are at positions $j=j_1$ and $j=j_2$. (b) The same finite element
model, now showing the temporally and spatially varying flux $\phi_{j}\left(t\right)$,
as well as the branch phase variables for the transmission line, $\varphi_{\text{TL},j}$,
and the ones describing the capacitive and Josephson coupling between
ground and transmission line, $\varphi_{j}$.\label{fig_finite_element_model}}
\end{figure}

To begin, we model the simple 1D version of the SQUID from Fig.~\ref{fig_intro}(a)
by means of a finite element approach, see Fig.~\ref{fig_finite_element_model}(a),
where the island is described by a transmission line. In terms of
the branch variables (see Fig.~\ref{fig_finite_element_model}(b)),
we find the Lagrangian
\begin{equation}
\mathcal{L}=\mathcal{L}_{\text{TL}}+\mathcal{L}_{V}+\mathcal{L}_{J}\label{eq:Lagrangian_total}
\end{equation}
with a part describing the transmission line (i.e., the island)
\begin{equation}
\mathcal{L}_{\text{TL}}=-\frac{1}{2L}\sum_{j=1}^{J}\left(\frac{\varphi_{\text{TL},j}}{2e}\right)^{2}\label{eq:Lagrangian_TL}
\end{equation}
and the capacitive coupling to ground
\begin{equation}
\mathcal{L}_{V}=\frac{C}{2}\sum_{j=0}^{J}\left(\frac{\dot{\varphi}_{j}}{2e}\right)^{2}.\label{eq:Lagrangian_V}
\end{equation}
Finally, the Josephson junctions are included in
\begin{equation}
\mathcal{L}_{J}=E_{J1}\cos\left(\varphi_{j_{1}}\right)+E_{J2}\cos\left(\varphi_{j_{2}}\right),\label{eq:Lagrangian_J}
\end{equation}
where the first junction is at position $j=j_{1}$, and the second
one at position $j=j_{2}>j_{1}$. The spatial dependence of the magnetic
field, $B\left(x\right)$ is here taken into account by the flux distribution
$\phi_{j}\left(t\right)$ (for $j=1\ldots J$) (see Fig.~\ref{fig_finite_element_model}(b)).
Thus, the branch variables are subject to the constraints
\begin{equation}
\varphi_{j-1}+\varphi_{\text{TL},j}-\varphi_{j}=\phi_{j}\left(t\right).\label{eq:constraints}
\end{equation}
This gives rise to $J$ constraints. Given that there are $J+1$ branch
variables for the capacitances, $\varphi_{j}$, all but one of these
variables can be eliminated. The transmission line variables $\varphi_{\text{TL},j}$
remain free at this stage. This is equivalent to the prescription
advocated by \citet{You2019} to uniquely determine the Hamiltonian.

Importantly, it can now be shown (for details, see Appendix \ref{app:low_energy})
that there is a low-frequency regime for the time-dependent driving
of the $\phi_{j}$ for which the transmission line dynamics are irrelevant,
such that $\varphi_{\text{TL},j}\approx 0$. This corresponds to the
situation where the driving is sufficiently slow that the transmission
line can quasi-instantaneously follow the perturbation (the exact
condition will be discussed below). The resulting low-frequency Hamiltonian
can be obtained by means of a Schrieffer-Wolff transformation,
\begin{multline}
\widehat{H}_{\text{low}}\approx\frac{1}{2}\frac{\left(2e\right)^{2}}{C_{\text{tot}}}\widehat{n}^{2}-E_{J1}\cos\left(\widehat{\varphi}+\phi_{1,\text{irr}}\right)\\
-E_{J2}\cos\left(\widehat{\varphi}+\phi_{2,\text{irr}}\right),\label{eq:H_low_discrete}
\end{multline}
with
\begin{equation}
\phi_{k,\text{irr}}=\frac{1}{J+1}\sum_{j=1}^{J}f_{j}\left(t\right)-f_{j_{k}}\left(t\right),
\end{equation}
for the junction index $k=1,2$, and having defined the (discrete)
flux integral $f_{j}\left(t\right)=\sum_{k=1}^{j}\phi_{k}\left(t\right).$
The total capacitance is simply $C_{\text{tot}}=\left(J+1\right)C$.
Crucially, when approaching the continuum limit for the network (i.e.,
the distance between the lumps $dx\rightarrow0$)
\begin{equation}
\phi_{j}\left(t\right)=2\pi\frac{\Phi_{j}\left(t\right)}{\Phi_{Q}}\rightarrow2\pi\frac{DdxB\left(x,t\right)}{\Phi_{Q}}
\end{equation}
where $D$ is also here the distance between the transmission line
and the ground and $W=\left(J+1\right)dx$ is the width. Thus, we recover
the result from Sec.~\ref{subsec:explicit_solution_1D}, Eq.~(\ref{eq:phi_irr_1D}).
This demonstrates the equivalence of the approach of You et al. \citep{You2019}
and our gauge prescription for the vector potential, $\mathbf{A}_{\text{irr}}$,
as outlined in Sec. \ref{irrotational_gauge}.

Finally, let us comment on the regime of validity of $H_{\text{low}}$
in Eq. (\ref{eq:H_low_discrete}). As we detail in Appendix \ref{app:low_energy},
neglecting the internal dynamics of the island is justified for driving
frequencies $\omega$ satisfying
\begin{equation}
\omega<\omega_{0}=\frac{\pi}{\sqrt{WlC_{\text{tot}}}}
\end{equation}
where $l$ is the inductance density of the transmission line, such
that $L=l\,dx$. The frequency $\omega_{0}$ corresponds to the lowest
mode of the internal transmission line dynamics. Assuming that the kinetic part of the inductance is dominant, we estimate $l\sim m_{\text{e}}/\left(n_{\text{s}}e^{2}\mathcal{S}\right)$~\cite{Little_1967,Meservey_1969},
with $m_{\text{e}}$ as the electron mass and the Cooper pair density
$n_{\text{s}}\sim4\times10^{6}\mu m^{-3}$ \citep{Wang_2014}, and
assuming a transmission line cross section of $\mathcal{S}\sim10\text{nm}\times20\mu m$
as well as $W\sim200\mu m$, we find $\omega_{0}\sim20\text{GHz}$.
For a typical qubit frequency between $5$ and $10\text{GHz}$, it
seems plausible that neglecting the internal island dynamics is justified
even when an ac magnetic field would be used to drive bit
flips.
\smallskip

\section{Conclusions\label{sec:conclusions}}

We develop a general recipe to construct a Hamiltonian for realistic
quantum circuits driven by external electomagnetic fields which vary
in time. This construction invokes the notion of an irrotational gauge
for the time-dependent vector field, which corresponds to the Coulomb
gauge, with the additional boundary condition that the vector potential
should be orthogonal to the conductor surfaces. Based on this result,
we show, in the example of a simple SQUID geometry, that assigning
individual capacitances to each Josephson junction leads to negative
and potentially time-dependent capacitances. We discuss measurable
effects of such anomalous capacitances, such as a revised prediction
for the qubit relaxation rate, and a nonzero Berry phase. Finally,
we establish a connection between the here proposed vector potential
gauge for continuous circuit geometries and irrotational gauge developed
in Ref.~\citep{You2019} for discrete circuits. In doing so, we also
provide an estimate for the validity of the lumped-element approach, given a certain driving frequency.
\begin{acknowledgments}
We thank F. Hassler and G. Catelani for stimulating discussions. We gratefully acknowledge funding by the German Federal Ministry
of Education and Research within the funding program "Photonic Research
Germany" under contract number 13N14891, and within the funding program "Quantum Technologies - From Basic Research to the Market" (project GeQCoS), contract number
13N15685. D. P. D. thanks the
OpenSuperQ project (820363) of the EU Flagship on Quantum Technology, H2020-FETFLAG-2018-03, for support.
\end{acknowledgments}

\appendix
\begin{widetext}

\section{More general Hamiltonians arising from the continuum electrodynamics model}\label{app_T_and_general_gauge}

In Sec.~\ref{sec:general-geometries}, we showed the choice of gauge that leads to a Lagrangian and Hamiltonian satisfying the ``irrotational condition'' of You et al.~\cite{You2019}. Here we show another choice that leads to forms that resemble the more general Lagrangians/Hamiltonians that arise in the previous discrete circuit analysis. For this purpose however, we first need the correct definition for the kinetic energy term which enters the Lagrangian.

We start with the definition for the energy stored in a capacitor via the electric power~\cite{citeulike:9606973},
\begin{equation}
E_Q=\int^t dt'\, V I \ ,
\end{equation}
where the lower integration limit is not specified as it merely provides an irrelevant constant shift (which we disregard here). Through $I=\partial_t Q$, we may immediately substitute the time integral for an integral over charge,
\begin{equation}
E_Q=\int^Q dQ'\, V(Q')\ ,
\end{equation}
or, through partial integration, for an integral over voltage
\begin{equation}
E_Q=VQ-\int^V dV'\, Q(V') \equiv VQ-E_V \ .
\end{equation}
Identifying $Q$ as the canonically conjugate momentum, we see that $E_Q$ and $E_V$ are related through a Legendre transformation. We thus conclude that the kinetic energy in the Lagrangian is given by $T=E_V$ [see also Eq.~\eqref{eq_T_irr} in the main text], whereas $E_Q$ corresponds to the kinetic energy of the Hamiltonian. Importantly, for the irrotational gauge, where the energies are quadratic in both $V$ and $Q$, $E_Q=E_V$, and the above distinction is moot. However, for the gauges considered in this appendix, there will appear linear terms where this distinction is important, and the definition of the Lagrangian kinetic energy via the voltage integral is imperative.

To continue, we introduce a {\em variably shifted irrotational gauge}. This is a simple variant on the irrotational gauge introduced in the main text. We now divide the electric field into two parts as in Eq.~\eqref{eq_decomposition},
\begin{equation}
    \mathbf{E}=\mathbf{E}_Q'+\mathbf{E}_{\dot B}'.
\end{equation}
Here $\bf{E}_Q'$ satisfies the same conditions as $\bf{E}_Q$, except that Eq.~\eqref{eq_island_charge} is replaced by an shifted island charge:
\begin{equation}
\epsilon\oiint_{\mathcal{S}_{i}}d^{2}x\left(\mathbf{n}_{i}\cdot\mathbf{E}_{Q}'\right)=Q-Q_0(t)=Q-\dot f_0(t) \ .\label{offis}
\end{equation}
Note that this island charge function $Q_0(t)$ should not be confused with the offset gate charge $n_g$ discussed at the end of Sec.~\ref{irrotational_gauge}. $n_g$ is a real physical quantity, while $Q_0(t)$, merely parameterizing a new set of gauges, cannot appear in any physical observable.
Equation (\ref{offis}) emphasizes that the shifted charge function can be taken to be time dependent, and we will assume that this time dependence is expressed as the time derivative of a fixed function $f_0(t)$.
The solution for the electric field is thus of the form
\begin{equation}
\mathbf{E}_Q'=\left(Q-Q_0\right)\mathbf{e}\ .
\end{equation}
The other contribution to the field, $\bf{E}_{\dot B}'$, must likewise be changed; instead of corresponding to zero island charge, it will require the island charge
\begin{equation}
    \epsilon\oiint_{\mathcal{S}_{i}}d^{2}x\left(\mathbf{n}_{i}\cdot\mathbf{E}_{\dot{B}}'\right)=\dot f_0(t)\ ,
\end{equation}
resulting in
\begin{equation}
\mathbf{E}_{\dot{B}}'=\mathbf{F}[\dot{\mathbf{B}}]-\dot{f}_0\mathbf{e}\ ,
\end{equation}
and the resulting new vector potential
\begin{equation}\label{eq_A_irr_prime}
\mathbf{A}_\text{irr}'=-\mathbf{F}[\mathbf{B}]+f_0\mathbf{e}\ .
\end{equation}
In this gauge the constitutive equation~\eqref{eq_Q_vs_V_irr} is replaced by
\begin{equation}
    Q-Q_0(t)=C_{tot}V_\text{irr}',
\end{equation}
defining the island potential $V_\text{irr}'$ in the new gauge (it is again uniform in space across the island). The new kinetic energy in the Lagrangian is
\begin{equation}
    T_\text{irr}'=\frac{C_{tot}}{2}\left( \frac{{\dot\varphi}'}{2 e}+\frac{\dot f_0}{C_{tot}} \right)^2.
\end{equation}
This introduces the superconducting phase variable in the variably-shifted irrotational gauge, $\dot\varphi'$. After Legendre transformation, this leads to the full Hamiltonian
\begin{equation}
    {\widehat H}_\text{irr}'=\frac{(2e)^2}{2C_{tot}}\left({\widehat n}'-\frac{{\dot f}_0}{2e} \right)^2-\sum_k E_{Jk}\cos\left({\widehat\varphi}'+\phi_{k,\text{irr}}+\frac{2ef_{0}}{C_{\text{tot}}}\right) \ ,
\end{equation}
where the shift $2ef_0/C_\text{tot}$ of the phase in the Josephson term follows from the line integral of the new $A_\text{irr}'$ in Eq.~\eqref{eq_A_irr_prime}. Indeed, $\widehat{H}_\text{irr}=\widehat{U}\widehat{H}_\text{irr}'\widehat{U}^\dagger -i \widehat{U}\partial_t\widehat{U}^\dagger$, with $\widehat{U}=e^{-i2ef_0\widehat{n}/C_\text{tot}}$. Here we see the more general kinetic energy form, with both quadratic and linear number operators, as it occurs in Eqs.~(13-14) of \cite{You2019}.

\section{Derivation of electric field in parallel plate capacitor}\label{app:parallel_plate_derivations}

We here show how to arrive at the solution for $\mathbf{E}_{\dot{B}}$ in the parallel plate capacitor, given in Eqs.~(\ref{eq_E_upper_plate}-\ref{eq_tangential_E}) in the main text.
We commence with the solutions in the interior of the superconductors, $y<0$ and $y>D$. Given the external magnetic field from Eq.~\eqref{eq_B_field} in the main text, the interior field is
\[
\mathbf{B}_{\text{int}}\left(y<0\right)=\left(\begin{array}{c}
B_{x}\left(x,z\right)\\
0\\
B_{z}\left(x,z\right)
\end{array}\right)e^{\frac{y}{\lambda}}\quad\mathbf{B}_{\text{int}}\left(y>D\right)=\left(\begin{array}{c}
B_{x}\left(x,z\right)\\
0\\
B_{z}\left(x,z\right)
\end{array}\right)e^{-\frac{y-D}{\lambda}}.
\]
Let us focus for concreteness on $y>D$. Here, we find that the supercurrent
flowing inside the superconductor is of the form
\[
\mathbf{j}\left(y>D\right)=\frac{1}{\mu_{0}\lambda}\left(\begin{array}{c}
-B_{z}\left(x,z\right)\\
0\\
B_{x}\left(x,z\right)
\end{array}\right)e^{-\frac{y-D}{\lambda}},
\]
due to the Maxwell equation $\nabla\times\mathbf{B}_{\text{int}}=\mu_{0}\mathbf{j}$.
The $y$-component of the current vanishes due $\nabla\times\mathbf{B}_{\text{ext}}=0$.
This must be so since charges cannot leave the surface. We see that current
conservation demands $\partial_{z}B_{x}\left(x,z\right)=\partial_{x}B_{z}\left(x,z\right)$,
which is likewise satisfied due to $\nabla\times\mathbf{B}_{\text{ext}}=0$
for $0<y<D$. We note that it is useful to compactify the expression
for the current
\begin{equation}
\mathbf{j}\left(y>D\right)=-\frac{1}{\mu_{0}\lambda}\left(\mathbf{n}_{y}\times\mathbf{B}_{\text{ext}}\right)e^{-\frac{y-D}{\lambda}}.
\end{equation}
For the opposite plate, at $y<0$, we find a current in the reversed
direction
\begin{equation}
\mathbf{j}\left(y<0\right)=\frac{1}{\mu_{0}\lambda}\left(\mathbf{n}_{y}\times\mathbf{B}_{\text{ext}}\right)e^{\frac{y}{\lambda}}.
\end{equation}
In this compact form, we can show that $\nabla\cdot\mathbf{j}=0$
via $\nabla\cdot\left(\mathbf{n}_{y}\times\mathbf{B}_{\text{ext}}\right)=-\mathbf{n}_{y}\cdot\left(\nabla\times\mathbf{B}_{\text{ext}}\right)=0$. Let us note that in the main text, we treat the case of a quasi one-dimensional parallel plate capacitor, where we eventually approximate $B_x\approx 0$ and $B_z(x,z)\approx B_z(x)$. In this case, the current would be mostly flowing in $x$-direction. Note that the continuity equation for the current, $\nabla\cdot\mathbf{j}=0$, is (weakly) broken within the above approximation. In order to respect it exactly, we would
have to retain a small $z$-component for the current, which however turns out to be irrelevant for the field solutions of interest. 

In the case of a time-dependently driven external magnetic field $\mathbf{B}\rightarrow\mathbf{B}\left(t\right)$,
the screening Meissner currents become likewise parametrically time-dependent.
Associated to this time-dependent current, we receive an electric
field inside the superconductor as per the first London equation,
$\partial_{t}\mathbf{j}=\frac{1}{\mu_{0}\lambda^{2}}\mathbf{E}$,
which consequently decays likewise with the London penetration depth. We thus arrive at Eqs.~\eqref{eq_E_upper_plate} and~\eqref{eq_E_lower_plate} in the main text.

We now turn to the solution for $\mathbf{E}_{\dot{B}}$
in the exterior, $0<y<D$, which has to satisfy the requirements described in Sec.~\ref{irrotational_gauge}, in particular the boundary condition, Eq.~\eqref{eq_E_boundary}. We obtain the right-hand side of Eq.~\eqref{eq_E_boundary} from the previously computed solutions of the interior, given explicitly in Eqs.~\eqref{eq_E_upper_plate} and~\eqref{eq_E_lower_plate}. As already pointed out in the main text, this problem is most conveniently solved by separating $\mathbf{E}_{\dot{B}}$
into a longitudinal and tangential component
$\mathbf{E}_{\dot{B}}=\mathbf{E}_{L}+\mathbf{E}_{T}$, 
where the longitudinal component is defined such that it is perpendicular
to the surface, $\mathbf{n}_{y}\times\mathbf{E}_{L}=0$. The tangential part on the other hand has to satisfy $\mathbf{n}_{y}\cdot\mathbf{E}_{T}=0$, and
\begin{equation}\label{eq_boundary_for_E}
\left.\mathbf{E}_{T}\right|_{y=0,D}=\left.\mu_{0}\lambda^{2}\partial_{t}\mathbf{j}\right|_{y=0,D}.
\end{equation}
As for the condition $\nabla\times\mathbf{E}_{\dot{B}}=-\dot{\mathbf{B}}$,
it turns out that it is very convenient to decompose
it into two separate conditions for each field, with a proportionality
prefactor $\gamma$ that needs to be determined,
\begin{align}
\nabla\times\mathbf{E}_{\dot{B},L} & =-\left(1+\gamma\right)\dot{\mathbf{B}}\\
\nabla\times\mathbf{E}_{\dot{B},T} & =\gamma\dot{\mathbf{B}}.
\end{align}
For the longitudinal field the ansatz,
\[
\mathbf{E}_{\dot{B},L}=\left(\begin{array}{c}
0\\
E_{L,y}\\
0
\end{array}\right)
\]
yields two differential equations
\begin{equation}\label{eq_condition_for_E}
\nabla\times\left(\begin{array}{c}
0\\
E_{L,y}\\
0
\end{array}\right)=\left(\begin{array}{c}
-\partial_{z}E_{L,y}\\
0\\
\partial_{x}E_{L,y}
\end{array}\right)=-\left(1+\gamma\right)\left(\begin{array}{c}
\dot{B}_{x}\left(x,z\right)\\
0\\
\dot{B}_{z}\left(x,z\right)
\end{array}\right).
\end{equation}
In order to solve these equations, it will be extremely helpful to explicitly use the relationships of the two
components of the magnetic fields, stemming from $\nabla\times\mathbf{B}_{\text{ext}}=0$
and $\nabla\cdot\mathbf{B}_{\text{ext}}=0$, which, when written out, read
\begin{align}
\partial_{z}\dot{B}_{x}\left(x,z\right) & =\partial_{x}\dot{B}_{z}\left(x,z\right)\label{eq:rot}\\
\partial_{x}\dot{B}_{x}\left(x,z\right) & =-\partial_{z}\dot{B}_{z}\left(x,z\right).\label{eq:div}
\end{align}
By means of the identities, Eqs. (\ref{eq:rot}) and
(\ref{eq:div}), it can be shown that a valid solution is
given by
\begin{align*}
E_{L,y} & =-\frac{1+\gamma}{2}\int_{c}^{x}dx'\left[\dot{B}_{z}\left(x',z\right)+\dot{B}_{z}\left(x',c'\right)\right]\\
 & +\frac{1+\gamma}{2}\int_{c'}^{z}dz'\left[\dot{B}_{x}\left(x,z'\right)+\dot{B}_{x}\left(c,z'\right)\right],
\end{align*}
where the integration constants $c$ and $c'$ will have to be fixed. First, let us show that the above proposed solution indeed solves Eq.~\eqref{eq_condition_for_E}. For the $z$-component, we get
\begin{align*}
\partial_{x}E_{L,y} & =-\frac{1+\gamma}{2}\left[\dot{B}_{z}\left(x,z\right)+\dot{B}_{z}\left(x,c'\right)\right]+\frac{1+\gamma}{2}\int_{c'}^{z}dz'\underbrace{\partial_{x}\dot{B}_{x}\left(x,z'\right)}_{\text{use }\nabla\cdot\mathbf{B}=0}\\
 & =-\frac{1+\gamma}{2}\left[\dot{B}_{z}\left(x,z\right)+\dot{B}_{z}\left(x,c'\right)\right]-\frac{1+\gamma}{2}\int_{c'}^{z}dz'\partial_{z'}\dot{B}_{z}\left(x,z'\right)\\
 & =-\frac{1+\gamma}{2}\left[\dot{B}_{z}\left(x,z\right)+\dot{B}_{z}\left(x,c'\right)\right]-\frac{1+\gamma}{2}\left[\dot{B}_{z}\left(x,z\right)-\dot{B}_{z}\left(x,c'\right)\right]\\
 & =-\left(1+\gamma\right)\dot{B}_{z}\left(x,z\right)
\end{align*}
and the same works for the $x$-component,
\begin{align*}
-\partial_{z}E_{L,y} & =\frac{1+\gamma}{2}\int_{c}^{x}dx'\left[\partial_{z}\dot{B}_{z}\left(x',z\right)\right]-\frac{1+\gamma}{2}\left[\dot{B}_{x}\left(x,z\right)+\dot{B}_{x}\left(c,z\right)\right]\\
 & =-\frac{1+\gamma}{2}\left[\dot{B}_{x}\left(x,z\right)-\dot{B}_{x}\left(c,z\right)\right]-\frac{1+\gamma}{2}\left[\dot{B}_{x}\left(x,z\right)+\dot{B}_{x}\left(c,z\right)\right]\\
 & =-\left(1+\gamma\right)\dot{B}_{x}\left(x,z\right)\ .
\end{align*}
We now need to fix the integration constants such that the surface
charge associated to the jump in the longitudinal component integrates
to zero, according to the discussion in Sec.~\ref{irrotational_gauge}. That is,
\begin{equation}
\int_{0}^{W}dx\int_{0}^{H}dzE_{L,y}\left(x,z\right)=0\ .\label{eq:charge_zero}
\end{equation}
We see that the following method works. We may exploit the additivity of different solutions to the differential equation. By replacing in the ansatz $c\rightarrow x''$
and $c'\rightarrow z''$, and integrating $x''$ ($z''$) from $0$ to $W$ ($H$) and dividing by $W$ ($H$), we arrive at Eq.~\eqref{eq_E_longitudinal} in the main text. The condition in Eq.~\eqref{eq:charge_zero} is satisfied, as one can convince oneself when following the instructions in the main text, given after Eq.~\eqref{eq_tangential_E}.

We are left with the tangential field $\mathbf{E}_T$. Here, we
have to satisfy the boundary conditions as given above, in Eq.~\eqref{eq_boundary_for_E}. As a reminder,
this field should fulfill both $\nabla\times\mathbf{E}_{\dot{B},T}=\gamma\dot{\mathbf{B}}$
and $\nabla\cdot\mathbf{E}_{\dot{B},T}=0$. We find that 
\begin{equation}
\mathbf{E}_{\dot{B},T}=\lambda\left(\begin{array}{c}
\dot{B}_{z}\left(x,z\right)\\
0\\
-\dot{B}_{x}\left(x,z\right)
\end{array}\right)\frac{D-2y}{D},
\end{equation}
which is equivalent to Eq.~\eqref{eq_tangential_E} in the main text, succeeds in satisfying both conditions. For the divergence, we find
\[
\nabla\cdot\mathbf{E}_{\dot{B},T}=\lambda\left(\partial_{x}\dot{B}_{z}\left(x,z\right)-\partial_{z}\dot{B}_{x}\left(x,z\right)\right)\frac{D-2y}{D}=0,
\]
again due to $\nabla\times\mathbf{B}_{\text{ext}}=0$. The curl condition results in
\[
\nabla\times\mathbf{E}_{\dot{B},T}=2\frac{\lambda}{D}\left(\begin{array}{c}
\dot{B}_{x}\left(x,z\right)\\
0\\
\dot{B}_{z}\left(x,z\right)
\end{array}\right)=2\frac{\lambda}{D}\dot{\mathbf{B}}\ .
\]
With this result, we finally identify $\gamma=2\lambda/D$
as given also in the main text.

\section{Berry curvature}\label{app:Berry_curvature}

Here, we show how to arrive at the approximate expression for the
Berry curvature in Eq. (\ref{eq:Berry_curvature}) in the main text.
Within the assumptions in the main text, in particular that $n_{g}$
is close to $1/2$, we write the Hamiltonian given in Eq. (\ref{eq:H_simple_case})
in the sub-basis of either one or zero extra Cooper pair on the island,
$\left|1\right\rangle ,\left|0\right\rangle $

\begin{equation}
\widehat{H}_{\text{low}}\approx\frac{1}{2}E_{C}\left(\left|1\right\rangle \left\langle 1\right|-\left|0\right\rangle \left\langle 0\right|\right)\delta n_{g}-\frac{1}{2}E_{J}\left(\phi\right)\left[\left|1\right\rangle \left\langle 0\right|e^{-i\frac{\delta\phi_{B}}{2}}+\left|0\right\rangle \left\langle 1\right|e^{i\frac{\delta\phi_{B}}{2}}\right]
\end{equation}
with $\delta n_{g}=n_{g}-1/2$. This Hamiltonian has the eigenvalues
$\pm\frac{1}{2}\sqrt{E_{C}^{2}\delta n_{g}^{2}+E_{J}^{2}\left(\phi\right)}$
and the corresponding eigenvectors
\begin{equation}
\left|\pm\right\rangle =\frac{1}{\sqrt{2}}\left(\sqrt{1\pm\frac{E_{C}\delta n_{g}}{\sqrt{E_{C}^{2}\delta n_{g}^{2}+E_{J}^{2}\left(\phi\right)}}}\left|1\right\rangle \mp\sqrt{1\mp\frac{E_{C}\delta n_{g}}{\sqrt{E_{C}^{2}\delta n_{g}^{2}+E_{J}^{2}\left(\phi\right)}}}e^{i\frac{\delta\phi_{B}}{2}}\left|0\right\rangle \right).
\end{equation}
Denoting $\left|-\right\rangle $ as the even parity ground state
$m=0$ in the main text, the above eigenvector can be plugged into
Eq. (\ref{eq:Berry_curvature}).

\section{Low-energy approximation\label{app:low_energy}}

Here we detail the steps required to arrive at the low-frequency
Hamiltonian, Eq.~(\ref{eq:H_low_discrete}), starting from the transmission
line model in Fig.~\ref{fig_finite_element_model}, described by the
Lagrangian in Eqs.~(\ref{eq:Lagrangian_total}), (\ref{eq:Lagrangian_TL}),
(\ref{eq:Lagrangian_V}), and (\ref{eq:Lagrangian_J}). As pointed
out in the main text, the branch variables $\varphi_{j}$ are subject
to the constraints Eq.~(\ref{eq:constraints}), such that only one
of them remains free. For simplicity, let us keep $\varphi_{0}$ as
the free variable. Hence, we find for $j\geq1$
\begin{equation}
\varphi_{j}=\varphi_{0}+\sum_{k=1}^{j}\varphi_{\text{TL},k}-f_{j}\left(t\right),
\end{equation}
having defined the (discrete) flux integral $f_{j}\left(t\right)=\sum_{k=1}^{j}\phi_{k}\left(t\right).$
Now we compute the conjugate momenta
\begin{align}
2en_{\text{TL},j} & =2e\frac{\partial\mathcal{L}}{\partial\dot{\varphi}_{\text{TL},j}}\\
2en_{0} & =2e\frac{\partial\mathcal{L}}{\partial\dot{\varphi}_{0}}
\end{align}
which results in
\begin{equation}
\left(\begin{array}{c}
2e\vec{n}_{\text{TL}}\\
2en_{0}+C\sum_{k=1}^{J}\frac{\dot{f}_{k}\left(t\right)}{2e}
\end{array}\right)=C\left(\begin{array}{cc}
\mathbf{J} & \vec{J}\\
\vec{J}^{T} & J+1
\end{array}\right)\left(\begin{array}{c}
\frac{\vec{\varphi}_{\text{TL}}}{2e}\\
\frac{\varphi_{0}}{2e}
\end{array}\right),
\end{equation}
where we defined the vectors
\begin{align}
2e\left(\vec{n}_{\text{TL}}\right)_{j} & =2en_{\text{TL},j}+C\sum_{k=j}^{J}\frac{\dot{f}_{k}\left(t\right)}{2e},\,\,\,1\leq j\leq J,\\
\left(\frac{\vec{\varphi}_{\text{TL}}}{2e}\right)_{j} & =\frac{\varphi_{\text{TL},j}}{2e}\\
\left(\vec{J}\right)_{j} & =J+1-j
\end{align}
and the matrix
\begin{equation}
\left(\mathbf{J}\right)_{jj'}=J+1-\text{sup}\left\{ j,j'\right\} .
\end{equation}
Formally, we thus find the Hamiltonian via a standard Legendre transformation
\begin{align}
H & =\frac{1}{2C}\left(\begin{array}{c}
2e\vec{n}_{\text{TL}}\\
2en_{0}+C\sum_{k=1}^{J}\frac{\dot{f}_{k}\left(t\right)}{2e}
\end{array}\right)^{T}\left(\begin{array}{cc}
\mathbf{J} & \vec{J}\\
\vec{J}^{T} & J+1
\end{array}\right)^{-1}\left(\begin{array}{c}
2e\vec{n}_{\text{TL}}\\
2en_{0}+C\sum_{k=1}^{J}\frac{\dot{f}_{k}\left(t\right)}{2e}
\end{array}\right)+\frac{1}{2L}\left(\frac{\vec{\varphi}_{\text{TL}}}{2e}\right)^{T}\left(\frac{\vec{\varphi}_{\text{TL}}}{2e}\right)\nonumber \\
 & -E_{J1}\cos\left(\varphi_{0}+\sum_{j=1}^{j_1}\left[\varphi_{u,j}+\varphi_{l,j}\right]-f_{j_1}\left(t\right)\right)-E_{J2}\cos\left(\varphi_{0}+\sum_{j=1}^{j_2}\left[\varphi_{u,j}+\varphi_{l,j}\right]-f_{j_2}\left(t\right)\right).
\end{align}
Due to the symmetric shape of the capacitance matrix, its inverse is of the form
\begin{equation}
\frac{1}{C}\left(\begin{array}{cc}
\mathbf{J} & \vec{J}\\
\vec{J}^{T} & J+1
\end{array}\right)^{-1}=\left(\begin{array}{cc}
\mathbf{C}^{\text{inv}} & \vec{C}^{\text{inv}}\\
\left(\vec{C}^{\text{inv}}\right)^{T} & C^{\text{inv}}
\end{array}\right)\ .
\end{equation}
In fact, it can be shown that
\begin{align}
\left(\mathbf{C}^{\text{inv}}\right)_{jj'} & =\frac{1}{C}\left(2\delta_{jj'}-\delta_{j,j'+1}-\delta_{j+1,j'}\right)\\
\left(\vec{C}^{\text{inv}}\right)_{j} & =-\frac{1}{C}\delta_{j1}\\
C^{\text{inv}} & =\frac{1}{C}.
\end{align}
We may thus decompose the Hamiltonian into
\begin{align}
H & =\frac{1}{2}\left(2e\vec{n}_{\text{TL}}\right)^{T}\mathbf{C}^{\text{inv}}\left(2e\vec{n}_{\text{TL}}\right)+\frac{1}{2L}\left(\frac{\vec{\varphi}_{\text{TL}}}{2e}\right)^{T}\left(\frac{\vec{\varphi}_{\text{TL}}}{2e}\right)\nonumber \\
 & +\left(2e\vec{n}_{\text{TL}}\right)^{T}\vec{C}^{\text{inv}}\left(2en_{0}+C_{0}\sum_{k=1}^{J}\frac{\dot{f}_{k}\left(t\right)}{2e}\right)\nonumber \\
 & +\frac{1}{2}C^{\text{inv}}\left(2en_{0}+C_{0}\sum_{k=1}^{J}\frac{\dot{f}_{k}\left(t\right)}{2e}\right)^{2}\nonumber \\
 & -E_{J1}\cos\left(\varphi_{0}+\sum_{j=1}^{j_1}\varphi_{\text{TL},j}-f_{j_1}\left(t\right)\right)-E_{J2}\cos\left(\varphi_{0}+\sum_{j=1}^{j_2}\varphi_{\text{TL},j}-f_{j_2}\left(t\right)\right),
\end{align}
All that is left is to find the eigendecomposition of $\mathbf{C}^{\text{inv}}$,
\begin{equation}
\mathbf{C}^{\text{inv}}=\sum_{q}\frac{1}{C_{q}}\vec{v}_{q}\vec{v}_{q}^{T}.
\end{equation}
We thus get
\begin{equation}
H=H_{0}+V
\end{equation}
with $H_{0}=H_{\text{SQUID}}+H_{\text{TL}}$ encompassing the separate
Hamiltonians for the SQUID and TL degrees of freedom
\begin{equation}
H_{\text{TL}}=\frac{1}{2}\sum_{q}\frac{\left(2e\right)^{2}}{C_{q}}n_{q}^{2}+\frac{1}{2L}\left(\frac{\varphi_{q}}{2e}\right)^{2}
\end{equation}
and
\begin{align}
H_{\text{SQUID}} & =\frac{1}{2}C^{\text{inv}}\left(2en_{0}+C_{0}\sum_{k=1}^{J}\frac{\dot{f}_{k}\left(t\right)}{2e}\right)^{2}\nonumber \\
 & -E_{J1}\cos\left(\varphi_{0}+\sum_{j=1}^{j_1}\varphi_{\text{TL},j}-f_{j_1}\left(t\right)\right)\nonumber \\
 & -E_{J2}\cos\left(\varphi_{0}+\sum_{j=1}^{j_2}\varphi_{\text{TL},j}-f_{j_2}\left(t\right)\right),
\end{align}
and the interaction term between the two
\begin{equation}
V=\sum_{q}\lambda_{q}\left(2en_{0}+C_{0}\sum_{k=1}^{J}\frac{\dot{f}_{k}\left(t\right)}{2e}\right)\left(2en_{q}\right).
\end{equation}
In above equations, we defined
\begin{align}
n_{q} & =\vec{v}_{q}^{T}\vec{n}_{\text{TL}}\\
\varphi_{q} & =\vec{v}_{q}^{T}\vec{\varphi}_{\text{TL}}\\
\lambda_{q} & =\vec{v}_{q}^{T}\vec{C}^{\text{inv}}.
\end{align}
The TL modes will have energies $\sqrt{1/\left(LC_{q}\right)}$, and
they couple to the SQUID degrees of freedom via the coupling parameter
$\lambda_{q}$. Charge phase quantization will transform the parameters
into operators, $\left[\widehat{\varphi}_{q},\widehat{n}_{q'}\right]=2ei\delta_{qq'}$,
$\left[\widehat{\varphi}_{0},\widehat{n}_{0}\right]=2ei$. The modes
$q$ may be diagonalised by the bosonic operators
\begin{equation}
\widehat{a}_{q}=\frac{1}{\sqrt{2}}\left(\frac{C_{q}}{L}\right)^{\frac{1}{4}}\left[\frac{\widehat{\varphi}_{q}}{2e}+i\sqrt{\frac{L}{C_{q}}}2e\widehat{n}_{q}\right].
\end{equation}
Let us now try to find a low-energy approximation for this Hamiltonian.
In order to do so, we need to evaluate it in the limit $L\rightarrow0$.
This will put the excitations of the TL modes to infinitely high energies,
and within the Josephson potential energy, we may set $\varphi_{q}\rightarrow0$,
such that the TLs will become lumped elements with one single superconducting
phase. The TL degrees of freedom can be eliminated by means of a Schrieffer-Wolf
transformation. We however have to be careful due to the time-dependence
in the Hamiltonian. 

Since we want to eliminate fast frequencies, we formally add the projector
$\widehat{P}$, which projects out all but the ground state of the
TL modes,
\begin{equation}
\widehat{P}=\mathbf{1}_{0}\otimes\left|0_{\text{TL}}\right\rangle \left\langle 0_{\text{TL}}\right|
\end{equation}
where $a_{q}\left|0_{\text{TL}}\right\rangle =0$ for all $q$. In
this separation of spaces, $H_{0}$ is purely block-diagonal
\begin{equation}
\widehat{H}_{0}=\widehat{P}\widehat{H}_{0}\widehat{P}+\left(1-\widehat{P}\right)\widehat{H}_{0}\left(1-\widehat{P}\right)
\end{equation}
whereas the interaction is purely off-diagonal
\begin{equation}
\widehat{V}=\left(1-\widehat{P}\right)\widehat{V}\widehat{P}+\widehat{P}\widehat{V}\left(1-\widehat{P}\right).
\end{equation}
According to the standard (time-independent) Schrieffer-Wolff transformation,
we have to find an anti-unitary operator $\widehat{S}$, such that
\begin{align}
\widehat{H}_{\text{low}} & =e^{\widehat{S}}\widehat{H}e^{-\widehat{S}}\\
 & \approx\widehat{H}_{0}+\frac{1}{2}\left[\widehat{S},\widehat{V}\right]+\mathcal{O}\left(V^{3}\right),
\end{align}
which requires
\begin{equation}
\left[\widehat{H}_{0},\widehat{S}\right]=\widehat{V}.
\end{equation}
There is an important caveat for time-dependent systems: here, the
transformation would in principle give rise to an extra term
\begin{align}
\widehat{U}\widehat{H}\widehat{U}^{\dagger}-i\widehat{U}\dot{\widehat{U}}^{\dagger} & \Rightarrow e^{\widehat{S}}\widehat{H}e^{-\widehat{S}}+i\int_{0}^{1}dre^{\widehat{S}r}\dot{\widehat{S}}e^{-\widehat{S}r}\\
 & \approx e^{\widehat{S}}\widehat{H}e^{-\widehat{S}}+i\dot{\widehat{S}}\quad\text{exactly equal if }\left[\widehat{S},\dot{\widehat{S}}\right]=0.
\end{align}
It can however be shown by means of the following scaling arguments, that this extra
term can be neglected. Namely, as initially stated, we want to take
the limit $L\rightarrow0$. This means that the high energies in $\left(1-\widehat{P}\right)\widehat{H}_{0}\left(1-\widehat{P}\right)$
diverge as $\sim L^{-1/2}$. At the same time, the interaction term
$\widehat{V}$ diverges with $\sim L^{-1/4}$ (see relationship between
$n_{q}$ and $a_{q}$). Consequently, $\widehat{S}$ has to scale
with $L^{1/4}$, which is therefore a small term. Now, for $\dot{S}$
to be important with respect to $\widehat{V}$ in the perturbation
series, the time-dependence should be on frequencies which scale at
least as large as $\sim L^{-1/2}$ - the same scaling as the TL eigenfrequencies.
As a consequence, as long as the time-dependent driving is on frequencies
slower than the TL eigenfrequencies, we can neglect this term. This
corresponds to an adiabatic approximation with respect to the time
scale of the TL dynamics. It is then justified to simply perform the
time-independent Schrieffer-Wolff transformation, with the time dependence appearing
parametrically. This can be thought of as an ``instantaneous'' adjustment
of the TL degrees of freedom to the time-dependent driving.

We can therefore continue by computing $\widehat{S}$, through the
same relationship as in the stationary case, $\left[\widehat{H}_{0},\widehat{S}\right]=\widehat{V}$,
leading to the standard solution for the low-frequency Hamiltonian
\begin{equation}
\widehat{H}_{\text{low}}\approx\widehat{P}\widehat{H}_{0}\widehat{P}-\widehat{P}\widehat{V}\left(1-\widehat{P}\right)\frac{1}{\widehat{H}_{0}}\widehat{V}\widehat{P}.
\end{equation}
Note that with $L\rightarrow0$, and due to the operators $\widehat{\varphi}_{q}\sim L^{1/4}$,
we can set $\widehat{\varphi}_{q}\rightarrow0$ within the Josephson
energies. Plugging in the explicit expressions for $H_{0}$ and $V$,
we arrive at
\begin{align}
\widehat{H}_{\text{low}} & \approx\frac{1}{2}\left(C^{\text{inv}}-\sum_{q}C_{q}\lambda_{q}^{2}\right)\left(2e\widehat{n}_{0}+C_{0}\sum_{j=1}^{J}\frac{\dot{f}_{j}\left(t\right)}{2e}\right)^{2}\nonumber \\
 & -E_{J1}\cos\left[\widehat{\varphi}_{0}-f_{j_1}\left(t\right)\right]-E_{J2}\cos\left[\widehat{\varphi}_{0}-f_{j_2}\left(t\right)\right].
\end{align}
In order to obtain the sought after result, we are left with evaluating
\begin{equation}
C^{\text{inv}}-\sum_{q}C_{q}\lambda_{q}^{2}=C^{\text{inv}}-\left(\vec{C}^{\text{inv}}\right)^{T}\left(\mathbf{C}^{\text{inv}}\right)^{-1}\vec{C}^{\text{inv}}.
\end{equation}
And in fact, from the identity
\begin{equation}
\left(\begin{array}{cc}
\mathbf{C}^{\text{inv}} & \vec{C}^{\text{inv}}\\
\left(\vec{C}^{\text{inv}}\right)^{T} & C^{\text{inv}}
\end{array}\right)C\left(\begin{array}{cc}
\mathbf{J} & \vec{J}\\
\vec{J}^{T} & J+1
\end{array}\right)=\left(\begin{array}{cc}
\mathbf{1} & 0\\
0 & 1
\end{array}\right)
\end{equation}
one can extract the relationships
\begin{align}
\mathbf{C}^{\text{inv}}\vec{J}+\vec{C}^{\text{inv}}\left(J+1\right) & =0\\
\left(\vec{C}^{\text{inv}}\right)^{T}\vec{J}+C^{\text{inv}}\left(J+1\right) & =\frac{1}{C}.
\end{align}
These identities are useful as follows. From the first identity it
follows that
\begin{equation}
\left(\vec{C}^{\text{inv}}\right)^{T}\vec{J}=-\left(J+1\right)\left(\vec{C}^{\text{inv}}\right)^{T}\left(\mathbf{C}^{\text{inv}}\right)^{-1}\vec{C}^{\text{inv}}
\end{equation}
This can be inserted into the second identity, and we find 
\begin{equation}
C^{\text{inv}}-\left(\vec{C}^{\text{inv}}\right)^{T}\left(\mathbf{C}^{\text{inv}}\right)^{-1}\vec{C}^{\text{inv}}=\frac{1}{\left(J+1\right)C},
\end{equation}
where we easily identify $\left(J+1\right)C_{0}=C_{\text{tot}}$ as
the total capacitance of the SQUID. Consequently, we find the effective
Hamiltonian
\begin{align}
\widehat{H}_{\text{low}} & \approx\frac{1}{2}\frac{1}{C_{\text{tot}}}\left(2e\widehat{n}_{0}+C_{0}\sum_{j=1}^{J}\frac{\dot{f}_{j}\left(t\right)}{2e}\right)^{2}\nonumber \\
 & -E_{J1}\cos\left[\widehat{\varphi}_{0}-f_{j_1}\left(t\right)\right]-E_{J2}\cos\left[\widehat{\varphi}_{0}-f_{j_2}\left(t\right)\right].
\end{align}
or when discarding the constant term
\begin{align}
\widehat{H}_{\text{low}} & \approx\frac{1}{2}\frac{\left(2e\right)^{2}}{C_{\text{tot}}}\widehat{n}_{0}^{2}+\frac{1}{J+1}\sum_{j=1}^{J}\dot{f}_{j}\left(t\right)\widehat{n}_{0}\nonumber \\
 & -E_{J1}\cos\left[\widehat{\varphi}_{0}-f_{j_1}\left(t\right)\right]-E_{J2}\cos\left[\widehat{\varphi}_{0}-f_{j_2}\left(t\right)\right].\label{H_eff}
\end{align}
With a time-dependent unitary transformation, $U=e^{i \overline{f} \widehat{n}_0}$ where $\overline{f}=\sum_{j=1}^J f_j/(J+1)$, we arrive
at the Hamiltonian in the irrotational gauge given in Eq. (\ref{eq:H_low_discrete}),
where the remaining SQUID degree of freedom is renamed $\widehat{n}_{0}\rightarrow\widehat{n}$
and $\widehat{\varphi}_{0}\rightarrow\widehat{\varphi}$, as in the
main text.

\end{widetext}

\bibliography{bibliography}

\end{document}